\documentclass[namedreferences]{SolarPhysics}
\usepackage[optionalrh]{spr-sola-addons} 
\usepackage{graphicx}        
\usepackage{courier}         
\usepackage{color}           
\usepackage{url}             

\newcommand{\vect}[1]{{\bf #1}}

\newcommand{\ie}{{\it i.e.}}
\newcommand{\be}{\begin{equation}}
\newcommand{\ee}{\end{equation}}
\newcommand{\beq}{\begin{eqnarray}}
\newcommand{\eeq}{\end{eqnarray}}

\newcommand{\aeta}[3]{  #1, {\it Astron. Astrophys.}, {\bf  #2}, #3.}

\newcommand{\aspj}[3]{  #1, {\it Astrophys. J.}, {\bf  #2}, #3.}

\newcommand{\sph}[3]{   #1, {\it Solar Phys.}, {\bf  #2}, #3.}

\small

\begin{document}
\begin{article}

\begin{opening}

\title{Time-Distance analysis of the Emerging Active Region NOAA 10790}

\author{S. \surname{Zharkov}$^{1}$ \sep
        M.J.   \surname{Thompson}
        }

\runningauthor{ZHARKOV AND THOMPSON}
\runningtitle{Time--Distance Analysis of Emerging Active Region}

\institute{Department of Applied Mathematics,
          University of Sheffield 
          Hounsfield Road, Hicks  
          Building, Sheffield, S3 7RH, UK \\
          University of Sheffield, UK \\
          $^{1}$ \email{s.zharkov @ sheffield.ac.uk}
             }

\date{Received ; accepted }

\begin{abstract}

We investigate the emergence of Active Region NOAA 10790 by means of time--distance helioseismology. Shallow regions of increased sound speed at the location of increased magnetic activity are observed, with regions becoming deeper at the locations of sunspot pores. We also see a long-lasting region of decreased sound speed located underneath the region of the flux emergence, possibly relating
to a temperature perturbation due to magnetic quenching of eddy diffusivity, or to a dense flux tube.
We detect and track an object in the subsurface layers of the Sun 
characterised by increased sound speed which could be related to emerging 
magnetic flux and thus obtain a provisional estimate of the speed of 
emergence of around $1 \, {\rm km \ s^{-1}}$.

\end{abstract}
\end{opening}

\section{Introduction}
\label{sec:Intro}

Local helioseismology provides a tool for studying the emergence and
evolution of active regions and complexes of solar activity.
Several theoretical scenarios for the formation of active regions on the solar
surface have been suggested within the context of the
dynamo theory explaining the
formation of the ``sunspot belt'' at low latitudes.   Recently developed
``flux-transport'' dynamo models include the magnetic
flux transport by meridional circulation \cite{Wang91, Choudhuri95, Dikpati99, NC02, Kriv05}, flux transport by an
interface 
\cite{parker93, charb97}
and overshoot dynamo
with a positive radial shear beneath the surface 
\cite{rudiger95}, 
or by a distributed dynamo with a negative shear \cite{roberts72}. 

All of the models agree that the most favourable and promising place for
the $\alpha \Omega $ dynamo is the region which includes the deepest
layers of the convection zone, 
the convective overshoot layer and the tachocline.
Nevertheless, it is difficult to keep flux tubes with strong magnetic field
(B >100 G) in the interior for a long time against magnetic buoyancy \cite{parker79} and then
deliver them to the surface within the sunspot belt.

In a flux-transport model it is possible to calculate the velocities of
macroscopic diamagnetic transport as a function of depth in the 
convective zone by taking into account the magnetic advection processes.
\inlinecite{Kriv05} estimated that in the subsurface layers the transport
velocity increases from $0.5 {\rm \ km\ s^{-1}}$ at a depth of 30 Mm to around
$0.9 {\rm \ km\ s^{-1}}$ at 10 Mm depth. However, the way in which the flux
travels to the surface is not fully understood. There are ongoing arguments
in favour of each model, and they can only be tested by comparison
with the observations of active regions and sunspots on the surface and
their appearance in the solar interior by means of helioseismology.

Time--distance helioseismology  \cite{Duval93} is a fast-growing area 
of helioseismology.
The data used for time-distance inversions are the estimated wave travel-times between different points on the 
surface of the Sun. The travel-time estimates are obtained by cross-correlating and averaging 
the Doppler-velocity signal at different locations at the Sun's surface and then obtaining the cross-correlation function. 
The travel times are then computed from cross-correlations either via Gabor wavelet fitting \cite{Kosovichev} or 
by comparing with a reference cross-correlation function (see \opencite{GB2005} and references therein). Because of the stochastic nature
of solar oscillations, substantial spatial and temporal averaging of the data is required
to measure the frequencies and travel times accurately.

The properties of the solar interior are then determined by, first, establishing the relations
between travel-time variations and internal properties 
(variations of the sound speed, flow velocity, magnetic field). This is followed by the inversion of these relations, 
which are typically cast as linear integral equations, to infer the internal properties.

\inlinecite{Chang99} applied the method of acoustic imaging to study the emergence of NOAA 7978, and reported the detection of the signature 
of upward-moving
magnetic flux in the solar interior.
The first time-distance investigation of an emerging active region was by  \inlinecite{Kosovichev2000}, wherein 
an active region which emerged on
the solar disk in January 1998, was studied with SOHO/MDI for eight days, both before and after
its emergence at the surface. The results have shown a complicated structure of the emerging region in the
interior, and suggest that emerging flux ropes travel very quickly through the depth range of our
observations, estimating the speed of emergence at about $1.3 {\rm \ km\ s^{-1}}$.

In Section \ref{Sec:DataMethod} of this paper the method and the data used in this work are described. In Section \ref{Sec:Discussion} the results of 
sound-speed inversions are presented, together with the discussion. This is followed by conclusions in Section \ref{Sec:Conclusions}.

\section{Data and Method}
\label{Sec:DataMethod}
\subsection{Time--Distance Helioseismology}
\label{Sec:Method}
The raw Doppler signal at any location is composed of velocity signals occurring due to convective motions on the scales of granulation
and supergranulation, as well as the signal due to the superpositions of the wavepackets set up by resonant acoustic modes. 
The supergranular signal can largely be removed by  filtering out frequencies smaller than 1.5 mHz. However even after removing 
the low frequencies and frequencies above the acoustic cut-off of about 5.3 mHz, the oscillation signals due 
to the nature of sources and sinks of acoustic waves set up by turbulent convection remain stochastic. 

Due to the fact that acoustic waves with the same horizontal phase speed ($\omega / k$) travel the same horizontal distance 
($\Delta$) the Gaussian phase-speed filter is tuned to select waves with horizontal phase
speed ($v_i$) near the ray theory value corresponding to distance $\Delta_i$. We define a number of such filters ($F_i$) corresponding to different distances ($\Delta_i$) as follows:
$$
F_i(k, \omega; \Delta_i)=\exp [-(\frac{\omega}{k} - v_i)^2 / 2\delta v_i^2 ],
$$ 
where $\omega$ is the frequency and $k$ is the horizontal wavenumber in the data powerspectrum. Each filter is applied by pointwise multiplication of the Fourier transform $S(k_x, k_y, \omega)$ of the observed velocity data.

From the filtered data, we calculate point-to-annulus cross-correlation 
functions, $C({\bf r}, \Delta_i, t),$ for each skip-distance and each 
position/pixel of our data, where ${\bf r}$ is the position vector, 
$\Delta_i$ is the distance between two surface bounces and $t$ is 
time. The obtained set of cross-correlation functions is then averaged spatially over the quiet Sun to obtain an estimate for the reference cross-correlation function, $C^{\rm ref}(\Delta_i, t)$. This is used to measure one-way travel-time perturbations at each location following Gizon and Birch (\citeyear{GB02, GB04}, also \opencite{jensen}) approach:
\begin{equation}
\label{eq_GBtauDiscrete}
\tau_\pm=\frac{\sum_i \mp f(\pm t_i) \dot{C}^{\rm ref}(t_i)[C(t_i)-C^{\rm ref}(t_i)]}
	{\sum_i f(\pm t_i) [\dot{C}^{\rm ref}(t_i)]^2},
\end{equation}
where $f(t)$ is a window function designed to pick out the signal from the first bounce (\ie, direct arrival times). Here $\tau_+$ and $\tau_-$ correspond to wave travelling outwards and inwards from the measurement location respectively.
We define the mean travel times $\tau({\bf r}, \Delta)$ as 
 $\tau({\bf r}, \Delta)=(\tau_+({\bf r}, \Delta)+\tau_{-}({\bf r}, \Delta))/2$.
 
As a first approximation, perturbations $[\delta\tau({\bf r}, \Delta)]$ in these mean travel times are linearly related to the sound-speed perturbations $[\delta c]$ in the wave propagation region, 
\begin{eqnarray}
\delta \tau( \vect{r}, \Delta) = \int \int_S {\rm d}{\bf r'} 
\int_{-d}^0 {\rm d}z K({\bf r- r'}, z;\Delta)\frac{{\mathnormal \delta} c^2}{c^2}({\bf r'}, z),
\label{eq_TT_Kernel}
\end{eqnarray}
where $S$ is the area of the region, $d$ is its depth and $\delta\tau({\bf r}, \Delta)$ is defined as the difference between the measured travel time at a given location ${\bf r}$ on the solar surface and the average of the travel times in the quiet Sun. The sensitivity kernel for the relative squared sound-speed perturbation is given by $K$. 

For the forward problem we use wave-speed sensitivity kernels estimated using the Rytov approximation by \cite{jensen}. The Rytov and ray approximations give very similar results, with the former being perhaps more reliable for inferring deeper structures 
\cite{Cv04}.

The code for travel--time inversions is based on a multi-channel deconvolution algorithm (MCD) \cite{jensen98} enhanced by the addition of horizontal regularization. Following \inlinecite{jensen01} we discretized the Equation (\ref{eq_TT_Kernel}):
$
d_i=\delta \tau ({\it \vect{k}}, \Delta_i ), \ 
G_{ij}=K({\it \vect{k}}, z_j; \Delta_i ), \ 
m_j=\delta s({\it \vect{k}}, z_j).
$
Then for each horizontal wavevector ${\it \vect{k}}$ we find the vector ${\it \vect{m}}$ that solves
\begin{eqnarray}
{\rm min} \{ \Vert (\vect{d}-G \vect{m})  \Vert^2_2 +\epsilon(\vect{k})^2 \Vert L\vect{m} \Vert^2_2 \},
\end{eqnarray}
where $L$ is a regularisation operator. In this work we chose to apply more regularisation at larger depths, where acoustic wave travel times provide less information, by setting
$L={\rm diag}(c_{0, 1}, \ldots, c_{0, n}),$ where $c_{0, i}$ is the sound speed in the $i$-th layer of the reference model. 
We heavily regularise small horizontal scales \cite{Cv05} by taking $\epsilon(\vect{k})^2=\epsilon_0^2 (1+(|\vect{k}|/k_{\rm max})^2)^p$ with $p=200$ and $\epsilon_0^2=5\times10^3$.
We invert for 14 layers in depth located at (0.36, 1.17, 2.11,
					3.28, 4.74, 6.41,
					8.6, 11.22, 14.28,
					17.78, 21.79, 26.31,
					31.41, 36.95) Mm.	

The errors due to realisation noise are estimated by processing the 24 quiet-Sun datacubes, calculating noise covariance matrix in travel-time measurements  \cite{jensen01, GB04, Cv05}, and then evaluating noise in inversions according to \cite{jensen01}. 

\begin{center}
\begin{table} 
\begin{tabular}[b]{|r|c|c|c|c|} 
\hline 
$i$ & $\Delta_i$ & Skip-Distance range, Mm & $v_i, {\rm \ (km \ s^{-1})}$ & $\delta v_i, {\rm \ (km \ s^{-1})}$ \\
\hline 
1 & 5.83 & 3.553\,--\,8.107  &  14.21  & 2.63 \\  
2 & 7.29 & 5.013\,--\,9.567  &  15.40  & 2.63 \\ 
3 & 11.664 & 9.387\,--\,13.941 & 17.49 &  3.33 \\
4 & 16.038 & 13.761\,--\,18.315 & 25.82  & 3.86 \\ 
5 & 20.412 & 18.135\,--\,22.689 & 30.46  & 3.86 \\ 
6 & 24.786 & 22.509\,--\,27.063 & 35.46  & 5.25  \\
7 & 29.16 & 26.883\,--\,31.437  & 39.71  & 3.05  \\
8 & 33.66 & 31.383\,--\,35.937  & 43.29  & 3.15  \\
9 & 37.91 & 35.633\,--\,40.1870 &  43.29  & 3.15 \\ 
10 & 42.28 & 40.003\,--\,44.557 &  47.67 &  3.57 \\ 
11 & 46.66 & 44.383\,--\,48.937 &  57.16  & 3.78  \\
12 & 51.01 & 48.733\,--\,53.287 &  57.16  & 3.78  \\ 
\hline
\end{tabular}
\caption{Phase speed filter parameters}
\label{Tab_PhaseSpeed}
\end{table}
\end{center}

\subsection{Data}
\begin{figure}
\begin{center}
\includegraphics[width=5.5cm]{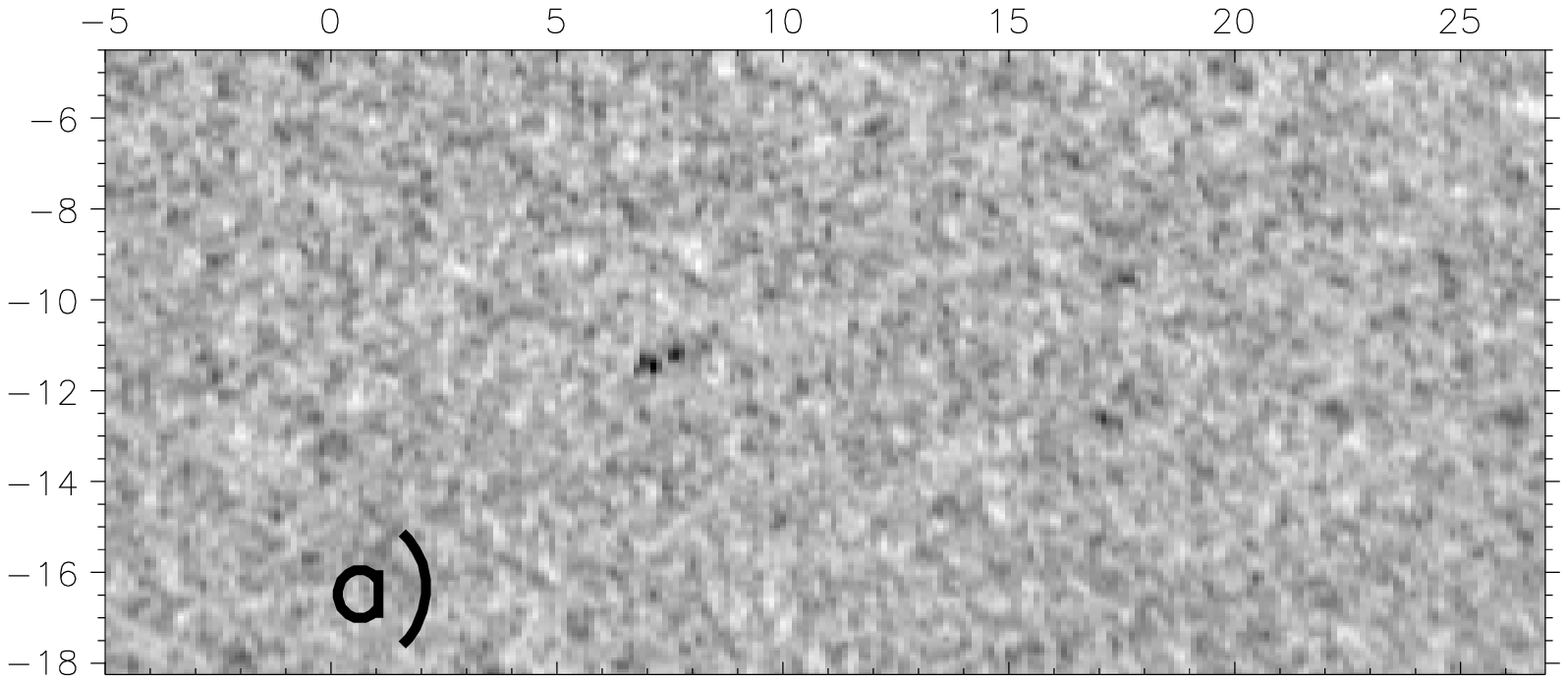} 
\includegraphics[width=5.6cm, height=2.4cm]{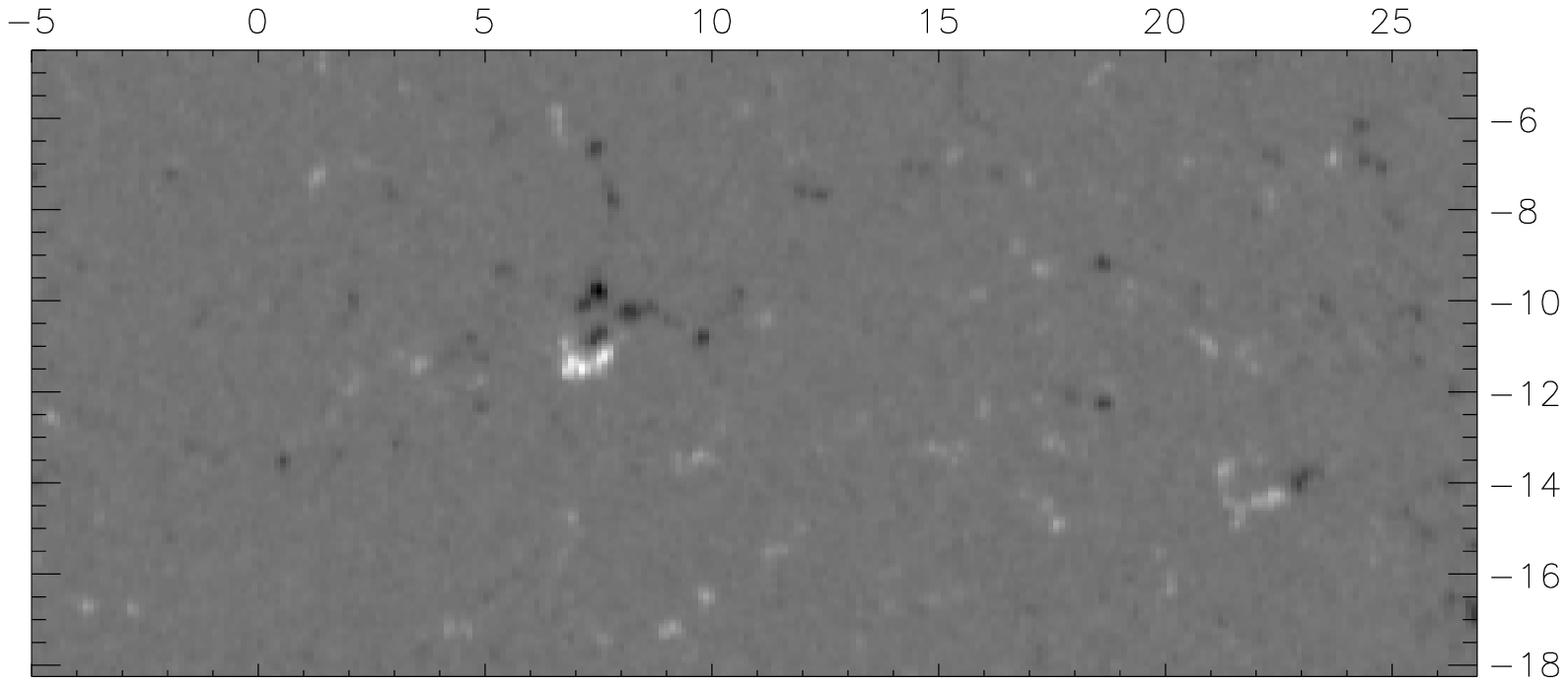} \\
\includegraphics[width=5.5cm]{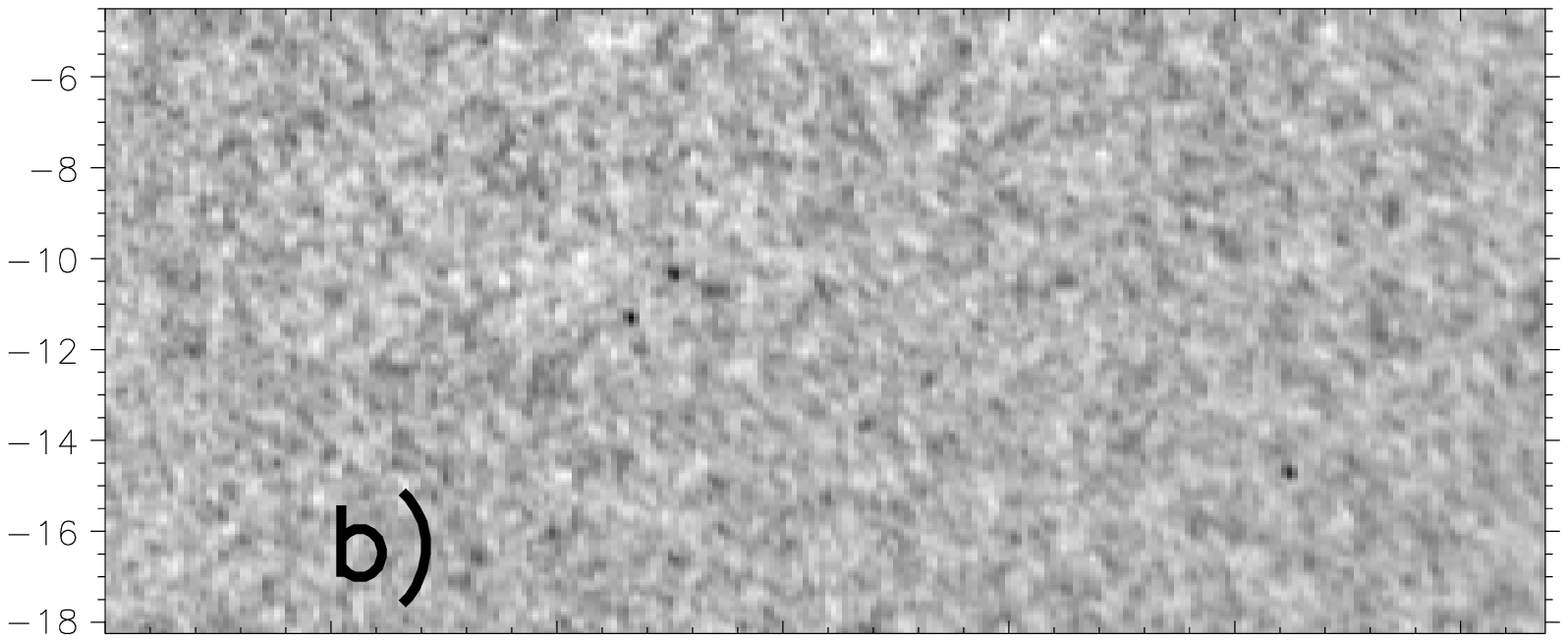} \ 
\includegraphics[width=5.5cm]{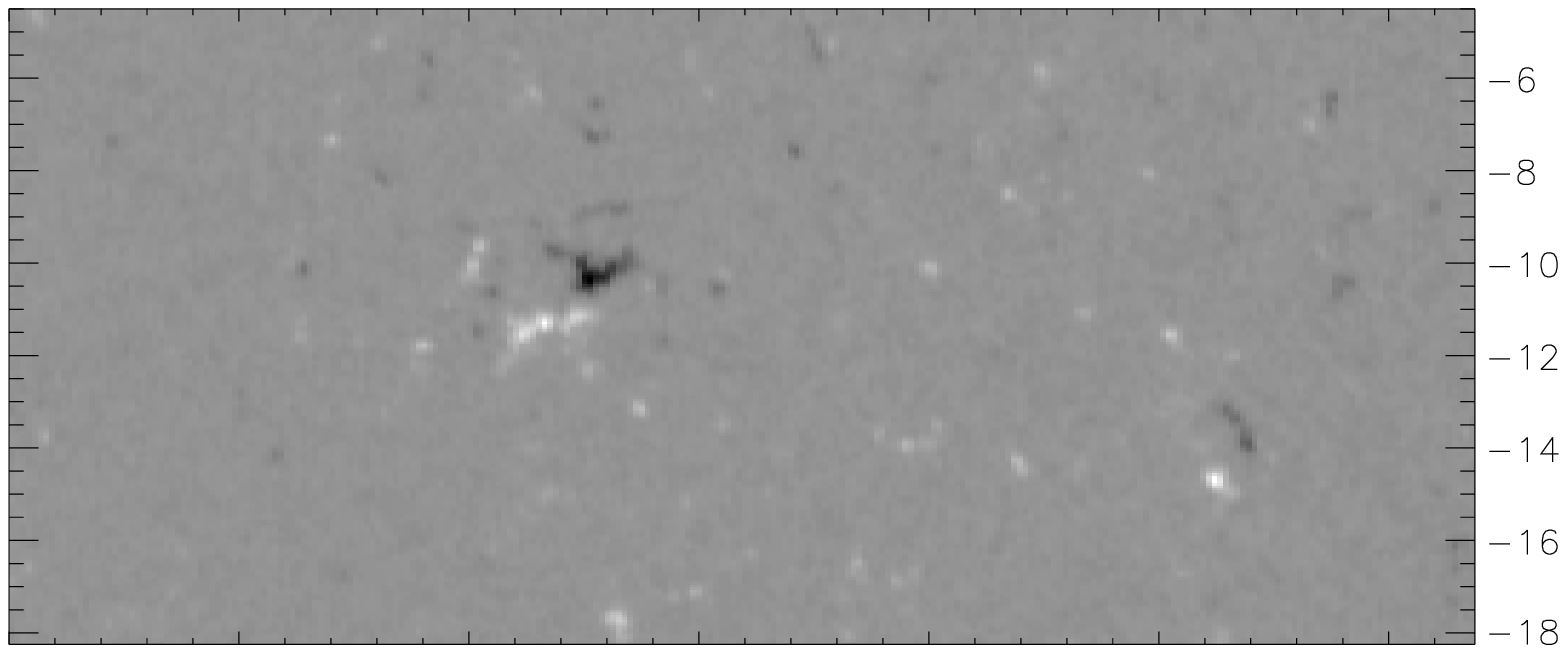} \\
\includegraphics[width=5.5cm]{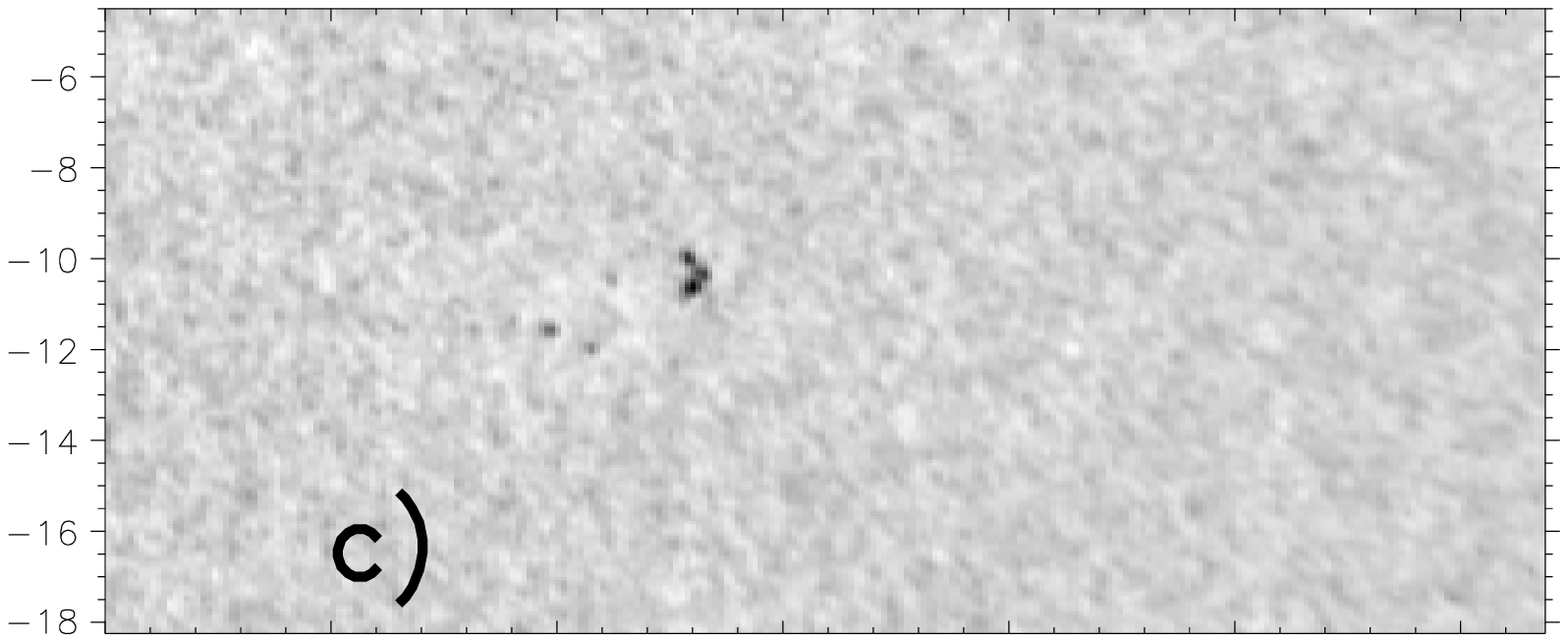} \ 
\includegraphics[width=5.5cm]{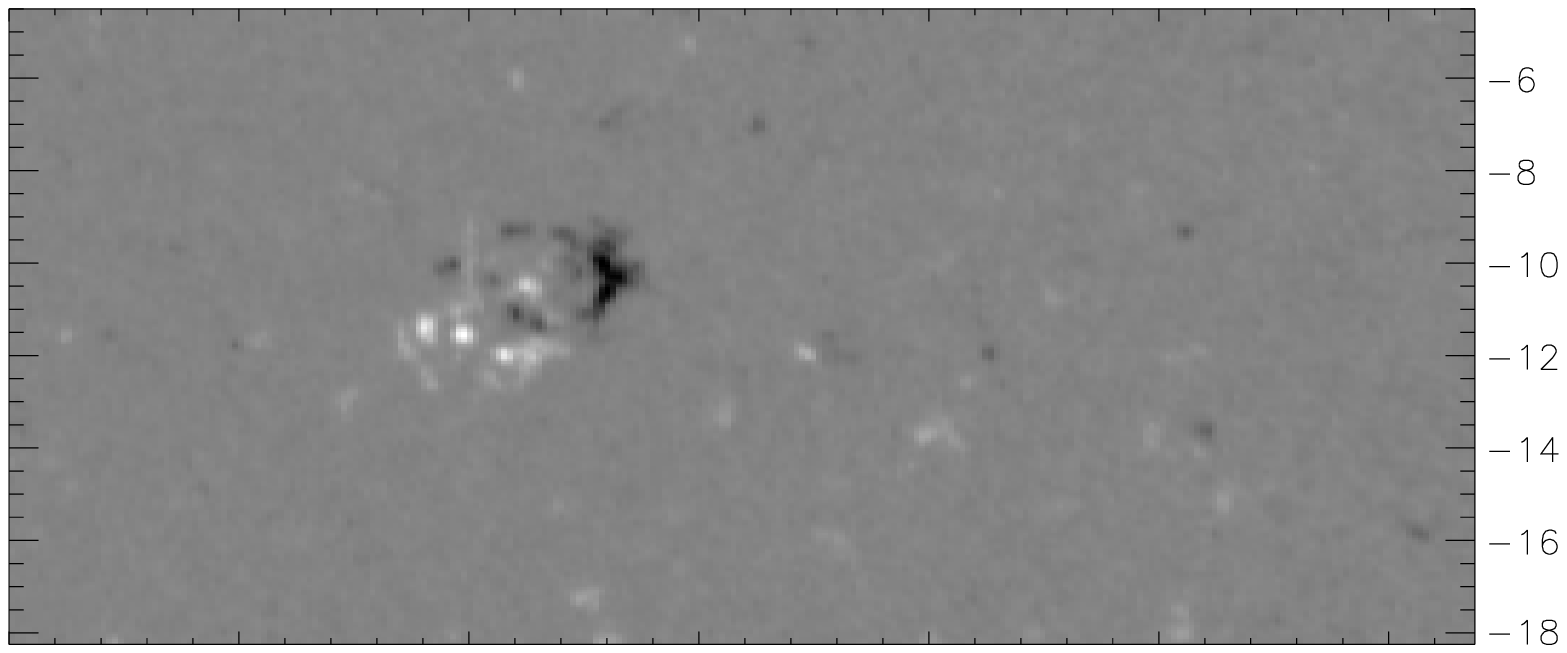} 
\\
\includegraphics[width=5.5cm]{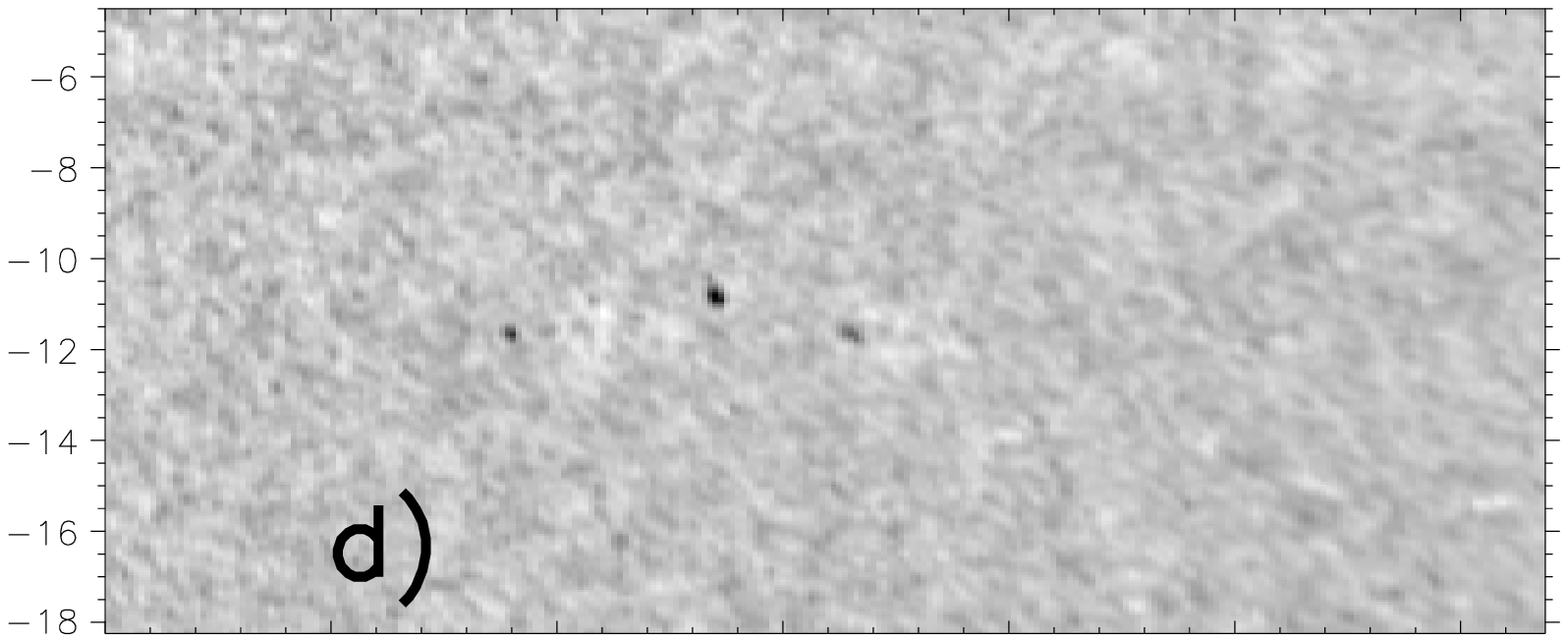} \ 
\includegraphics[width=5.5cm]{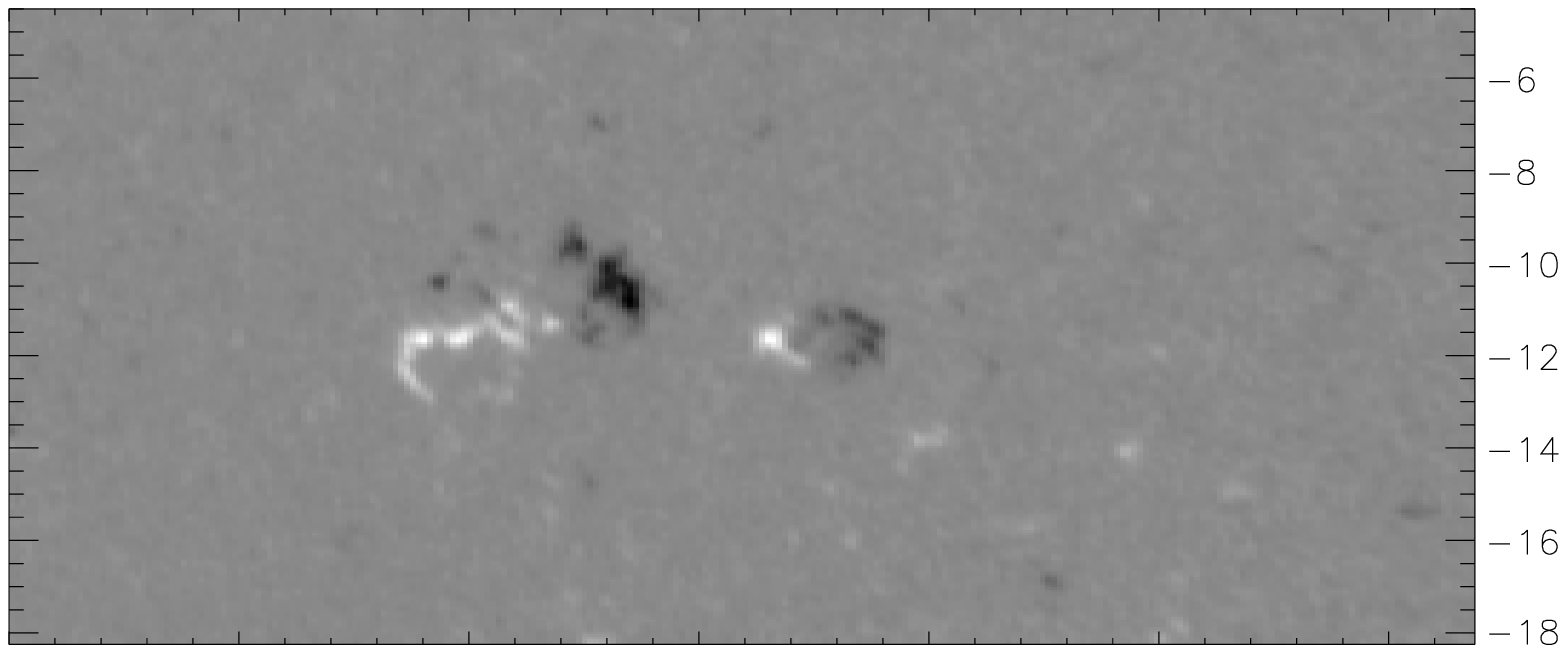}\\
\includegraphics[width=5.5cm]{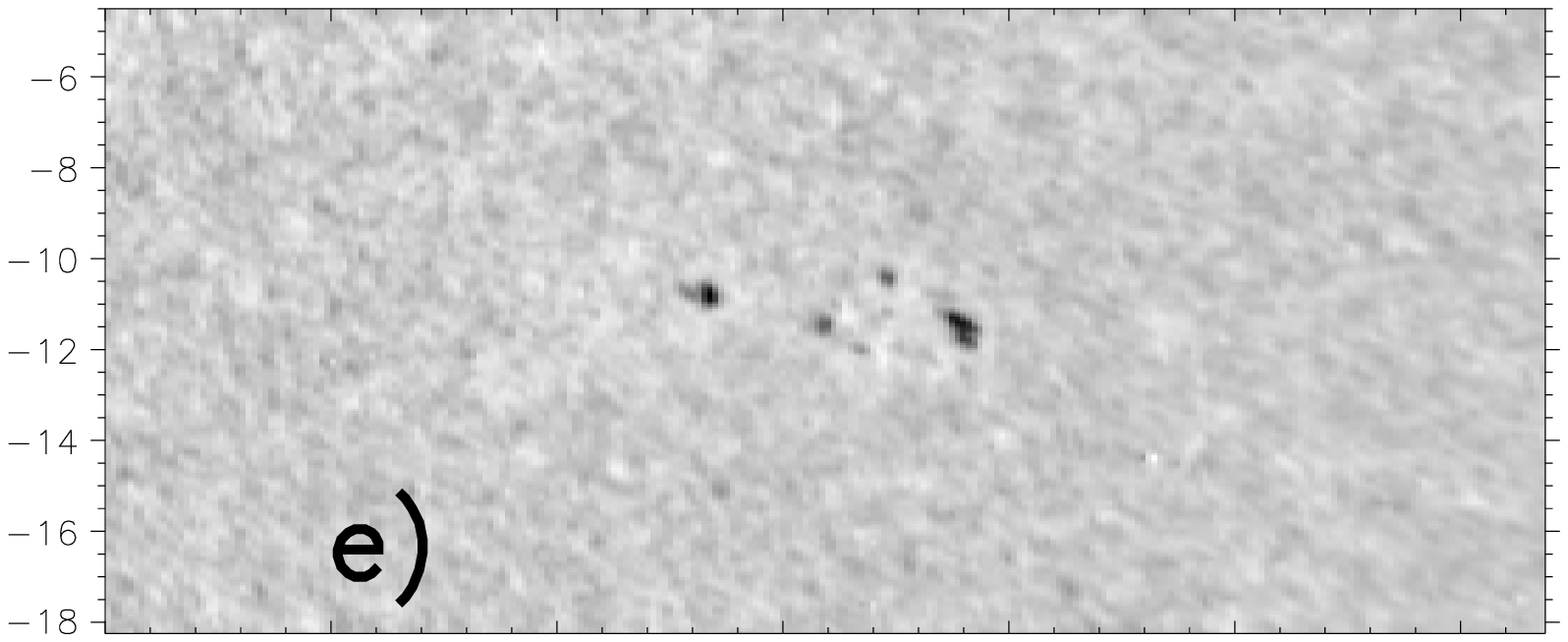} \  
\includegraphics[width=5.5cm]{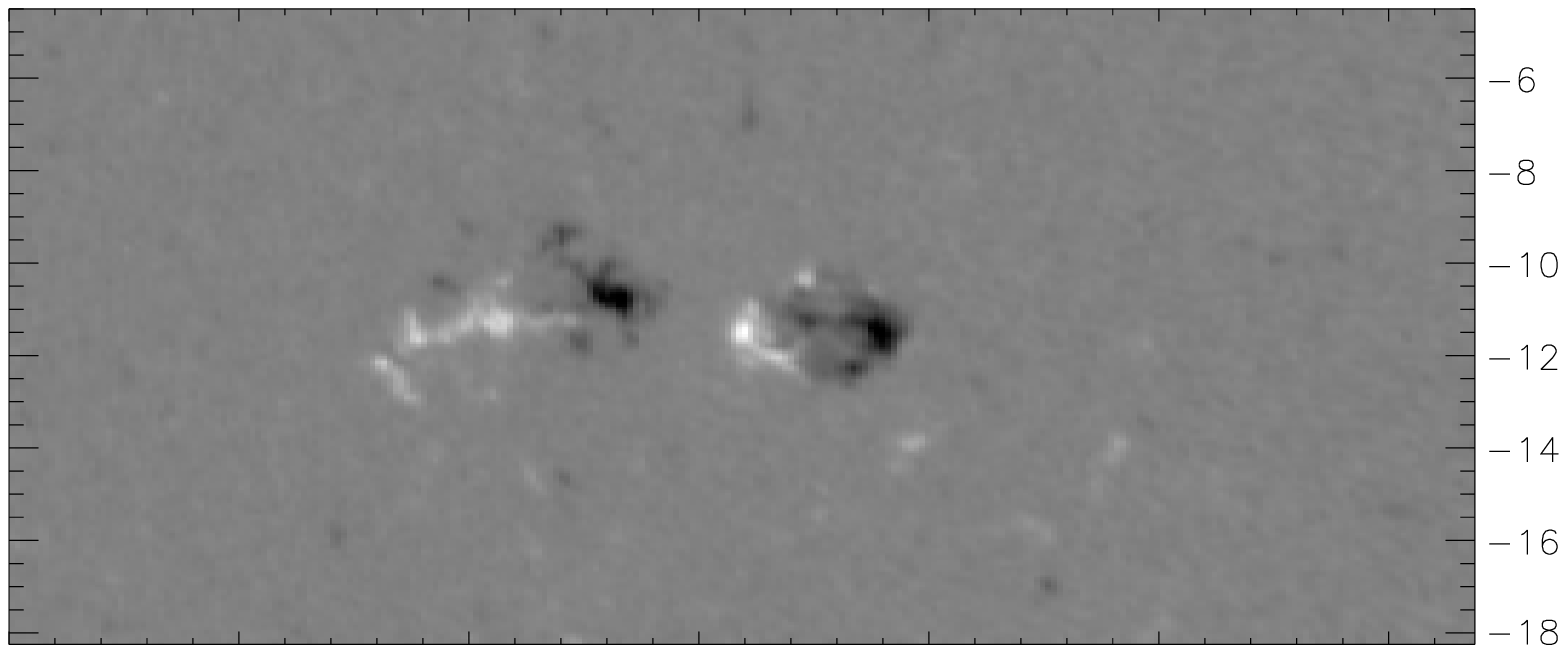} 
\\
\includegraphics[width=5.5cm]{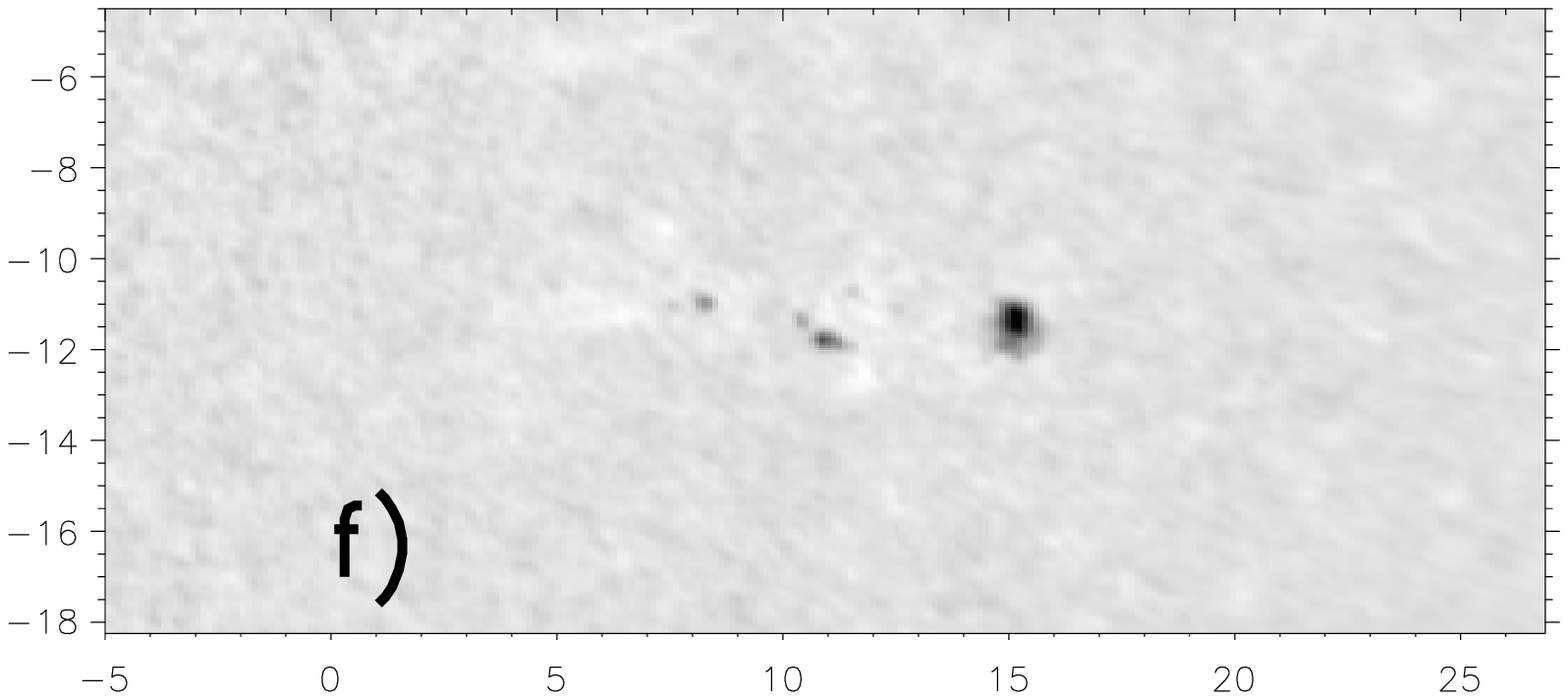}
\includegraphics[width=5.6cm, height=2.45cm]{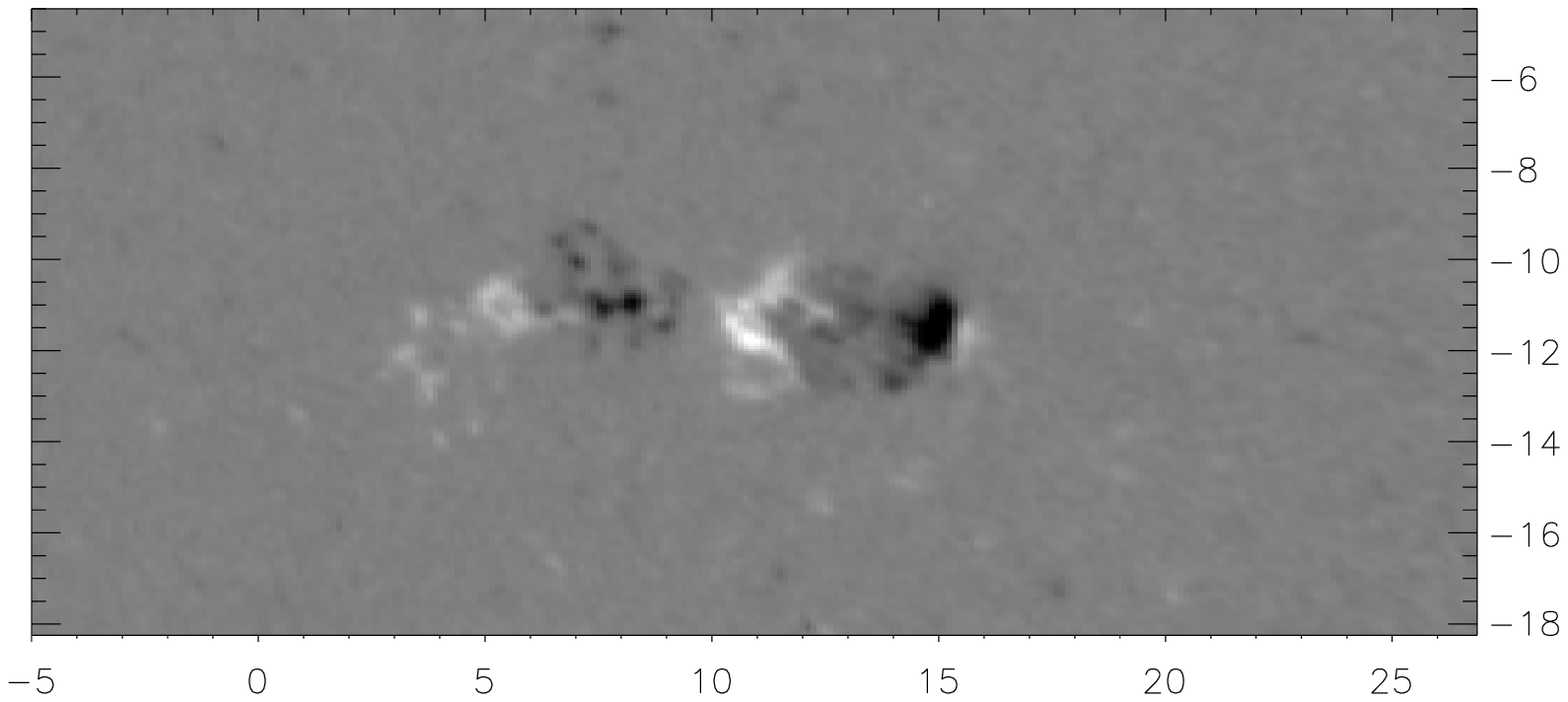} \\
\caption{The evolution of active region NOAA 10790  as observed by MDI. Carrington longitude is plotted along the $x$-axis, latitude along the $y$-axis. MDI continuum images are in the left column, the greyscale for MDI magnetograms {(right column)} corresponds to -600 to 600 Gauss. The  observation times for each row: a) 11 July 2005 11:11:32 UT, b) 12 July 2005 06:23:32UT, c) 13 July 2005 06:23:32 UT, d) 13 July 2005 17:35:32 UT, e) 13 July 2005 23:59:32 UT, f) 14 July 2005 11:11:32 UT.}
\label{fig:cnEvol}
\end{center}
\end{figure}

We investigate the active region 
NOAA 10790 appearing on 13 July 2005. 
The full disk MDI data with one minute cadence are used. The time series start at 00:00 UT, 10 July 2005,  and ends at 23:59 UT 13 July 2005. From this set we extract forty four 512-minute series with the starting times every two hours. Each dataset is then re-mapped and de-rotated onto the heliographic grid using a Postel projection centred at $11^\circ$ Carrington longitude and $11^\circ$ latitude (South). The horizontal spread of the data is approximately 380 Mm. We use the Snodgrass rotation rate to remove the differential rotation \cite{Snodgrass90} for each processed datacube. 

Each of the datacubes is then processed to obtain the travel-time measurements using the centre-to-annulus cross-correlations as described in Section \ref{Sec:Method}. The phase-speed filter values and the range of skip-distances used in this paper are provided in Table \ref{Tab_PhaseSpeed}. The averaging annuli thickness was set to three pixels wide which corresponds to approximately 4.55 Mm.

\section{Results and Discussion}
\label{Sec:Discussion}
\subsection{The Active Region NOAA10790 Surface Emergence and Evolution}
\label{sec:DisData}

Signs of the increased magnetic activity at the site of the active-region emergence can be first observed on SOHO/MDI magnetogram images taken on 10 July 2005. At this time one can see the two points of flux at around $5^\circ$\,--\,$8^\circ$ Carrington longitude. The flux slowly grows with the first pores appearing in the SOHO/MDI continuum images around 11 July 2005 06:30\,UT (Figure \ref{fig:cnEvol}). A number of different pores evolve and disappear at this location until 13 July 2005 06:30\,UT, but no sunspots develop. 

Also around 14:30\,UT 13 July 2005, we see a new flux emergence to the east of the first group at around $14^\circ$ longitude, $10$\,--\,$12^\circ$ latitude (see Figure \ref{fig:cnEvol}, panels d)\,--\,f)). At this new location, the first sunspot pores appear at about 17:30\,UT 13 July 2005, which quickly evolve into the bipolar sunspot group (by 14 July 2005 14:30\,UT) before disappearing behind the limb.  Also, there is a bipolar region of relatively weak magnetic flux seen appearing at around $20$\,--\,$22^\circ$  longitude on 10 July 2005  and 12 July 2005. No pores were observed at these locations.

\begin{figure}
\begin{center}
\includegraphics[width=5.7cm]{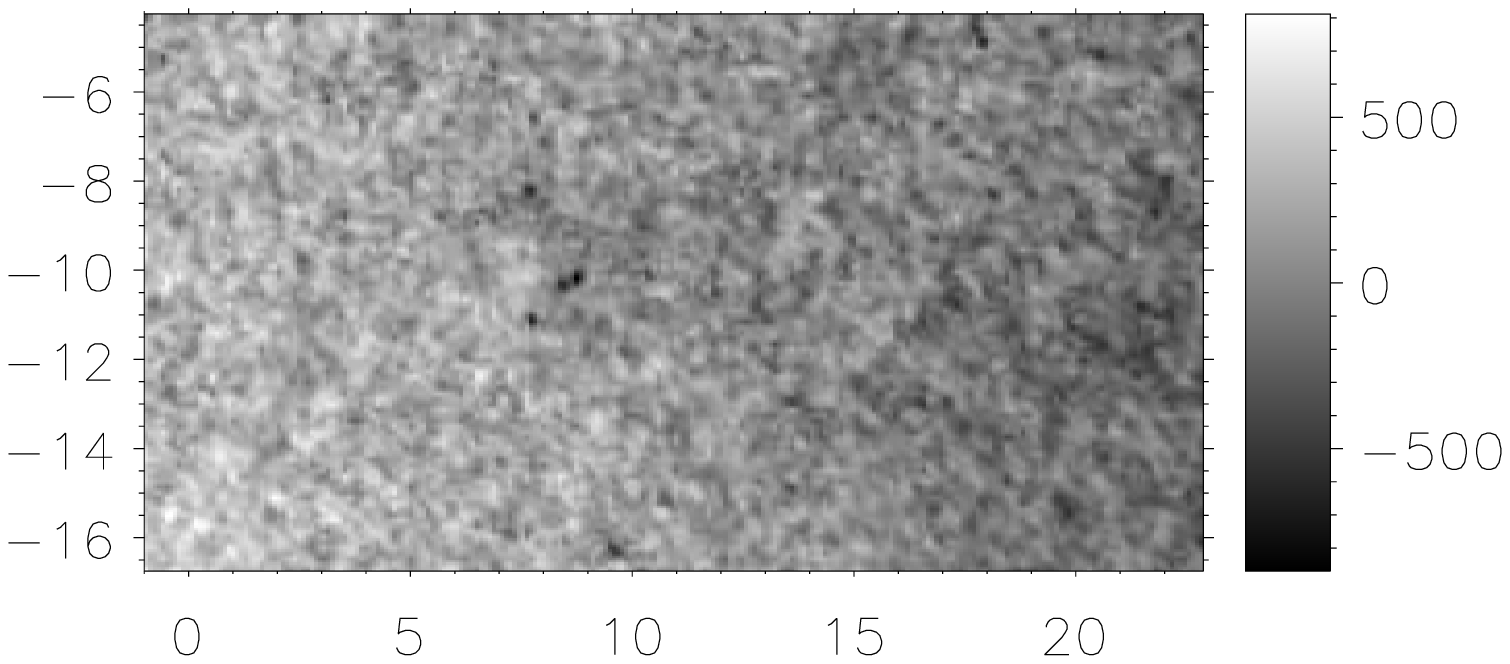} 
\includegraphics[width=5.7cm]{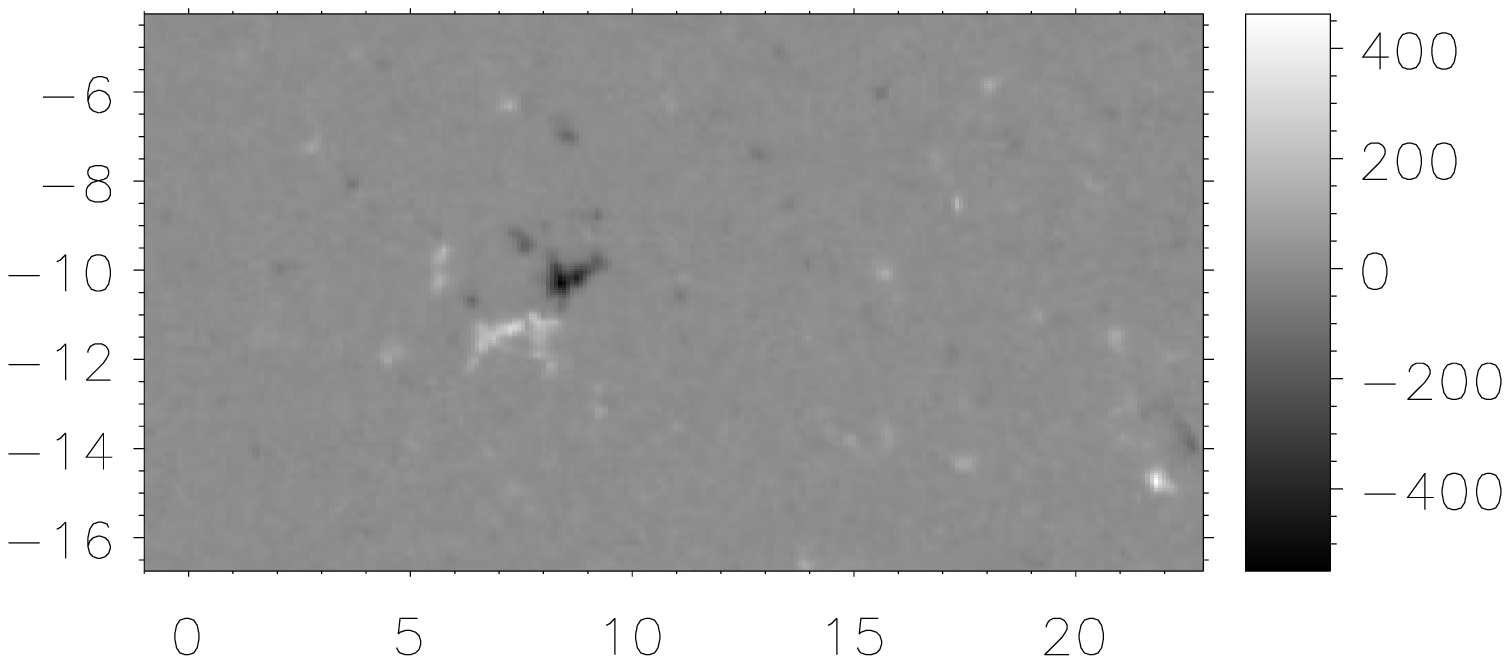} \\
\ \includegraphics[width=5.5cm]{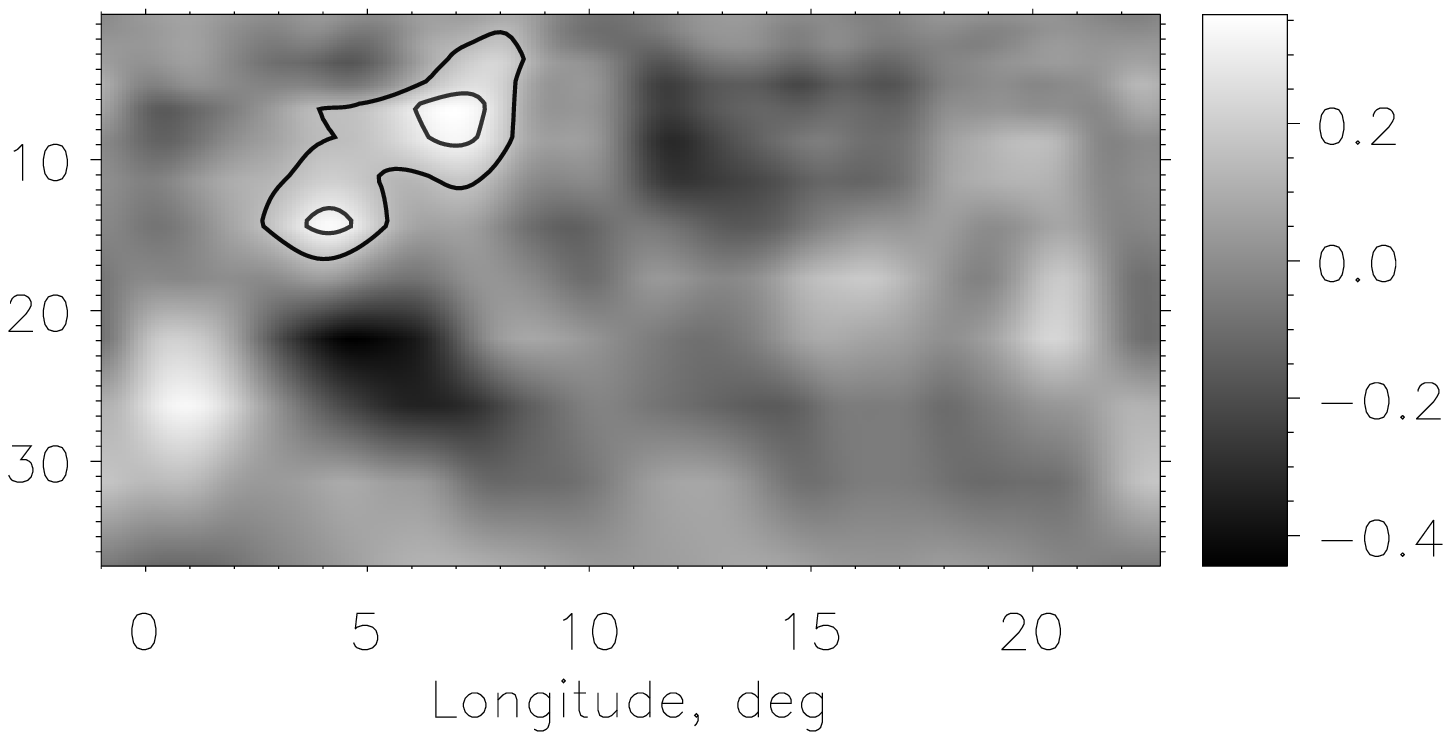}
\  \includegraphics[width=5.5cm]{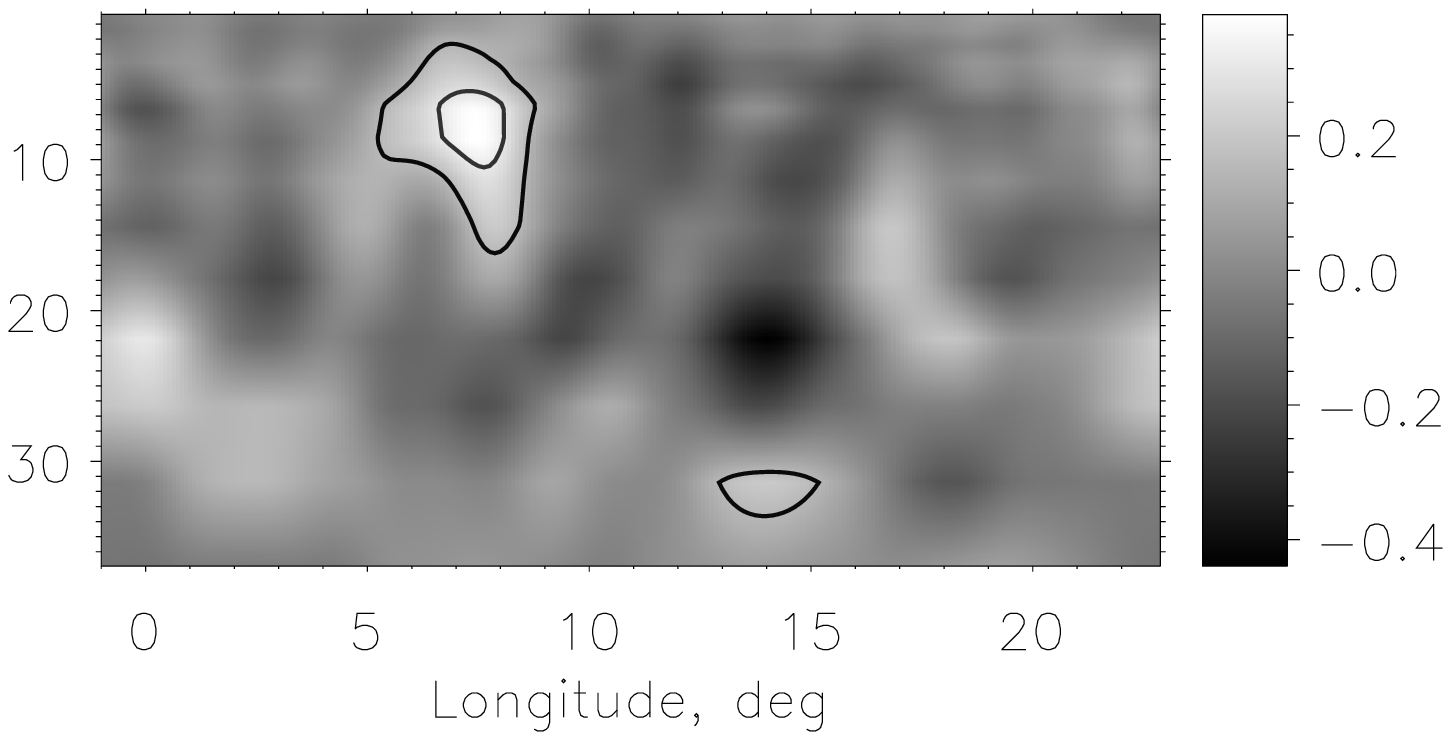}
\caption{Sunspot pore emergence. Top panels: the SOHO/MDI intensity (left) and magnetogram (right) images, taken on 12 July 2005 06:23\,UT and 09:35\,UT. Bottom panels: the cuts at 11 degrees of latitude through the sound-speed inversion estimated from 512 minute dopplergram timeseries centred on 11 July 2005 08:00\,UT ({left}) and 11 July 2005 10:00\,UT ({right}). The greyscale is in ${\rm km \ s^{-1}}$. Latitude is plotted on the left hand side along the $y$-axis and longitude along the $x$-axis. Depth in Mm is marked on the right hand side of $y$-axis. Contours correspond to sound-speed perturbation values of $0.25$ and $0.3$ ${\rm km \ s^{-1}}$.
}
\label{fig:fluxPore}
\end{center}
\end{figure}

\begin{figure}
\begin{center}
\includegraphics[width=12.cm, height=10.5cm]{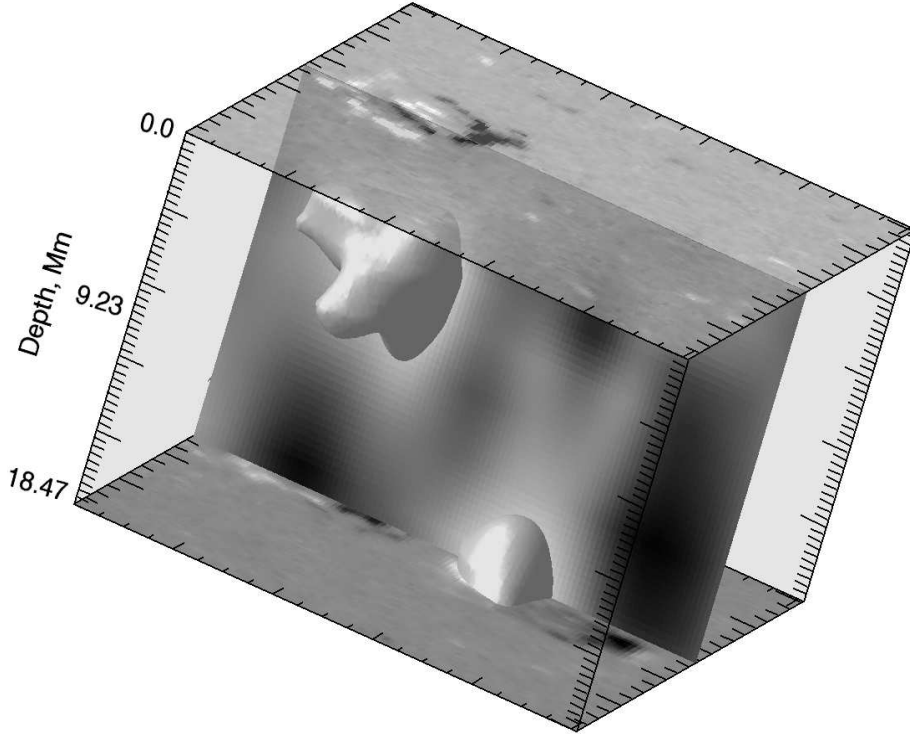}
\caption{The sound-speed inversion of a 512 minute datacube centred around 13 July 2005 04:00\,UT. The top is the surface magnetogram at the start of the time series. The bottom is the surface magnetogram taken around the time of the end of the whole series (14 July 2005 07:59\,UT). The vertical panel is the absolute sound-speed perturbation at $11^\circ$ latitude. The iso-surface corresponds to a value of 0.32 ${\rm km \ s^{-1}}$.}
\label{fig:inv3D}
\end{center}
\end{figure}

\subsection{The Subsurface Sound-Speed Inversions}

Inversions of the sound-speed perturbation beneath the region of interest carried out in this work indicate that little change is detected at the first site of increased magnetic activity ($6$\,--\,$10^\circ$ longitude) prior to the emergence 
of magnetic flux at around 11:00\,UT 10 July 2005. 
Then, once the magnetic flux appears on the surface, we see a consistent 
and substantial increase of the sound speed at shallow depths of up to 7\,--\,10 Mm directly beneath the locations of
high magnetic field strength, up until the end of the series used in this investigation. However, we have not been able to see any significant positive sound-speed perturbation in the deeper layers most of the time. 
This could be due to the fact that the spatial extent of the emerging flux at this location is quite 
small, six\,--\,seven Mm in diameter for each footprint, and the signal cannot be resolved to better than half of the wavelength \cite{GB04}. For
the power-spectrum produced by the phase-speed filters specified in Table \ref{Tab_PhaseSpeed}, this
 gives us an estimated resolution of, at best, $7.5$ Mm for waves travelling to $20$ Mm depths. 
The averaging kernel at a target depth of 18 Mm, estimated at full width at half-maximum, extends from 8 to 22.5 Mm.
This can also be a byproduct of the inversion regularisation, since due to our choice of operator, more regularisation is applied at deeper layers. 

As the ratio of gas pressure to magnetic pressure (the plasma $\beta$) increases rapidly with depth, the sound speed increase due to magnetic field becomes very small, $\delta c/c$ of order 0.01 per cent at around 18 Mm below the surface. Using the relations $p_e=k T_e \rho_e / m$, $p_i=k T_i \rho_i / m$, $p_m=\frac{B^2}{2\mu}$ and $p_e=p_i+p_m$, and assuming that the measured wave-propagation speed is given by $c^2=c_i^2+c_A^2$, where the Alfv\'en speed is $c_A^2=\frac{B^2}{\mu \rho_i}$, we can evaluate 
$$ \frac{\delta c^2}{c_e^2}= \frac{\delta T}{T_e} + \frac{2}{\gamma \beta}(1+\frac{\delta T}{T_e}),$$
Here $p$ is pressure, $T$ temperature, $\rho$ density, $k$ Boltzman constant, $m$ mean molecular weight, and indices $e, i, m$ stand correspond to {\it external}, {\it internal}, and {\it magnetic}; $\beta=p_i/p_m$, $\delta c^2=c^2-c_e^2,$ and $\delta T=T_i-T_e$. Thus, at greater depths the measured perturbation will be dominated by changes in temperature. This makes detecting and  following localised  perturbations due  to emerging  magnetic fields difficult.

Notwithstanding the argument above, in our measurements the positive perturbation below the surface regions of magnetic activity becomes deeper and better defined as sunspot pores begin to form at the surface, indicating perhaps the increase of magnetic field strength. 
In the two bottom plots of Figure \ref{fig:fluxPore} we can see the region of the increased sound speed extending to depths of around 12\,--\,15 Mm at around $7^\circ$ longitude that, we believe, is related to the appearance of a sunspot pore seen in SOHO/MDI intensity images. By the time of 13 July 2005 00:00\,UT there are several pores observed at this location, and the flux configuration has a clear bi-polar structure (Figure \ref{fig:cnEvol}c). 

\begin{figure}
\begin{center}
\includegraphics[width=6cm]{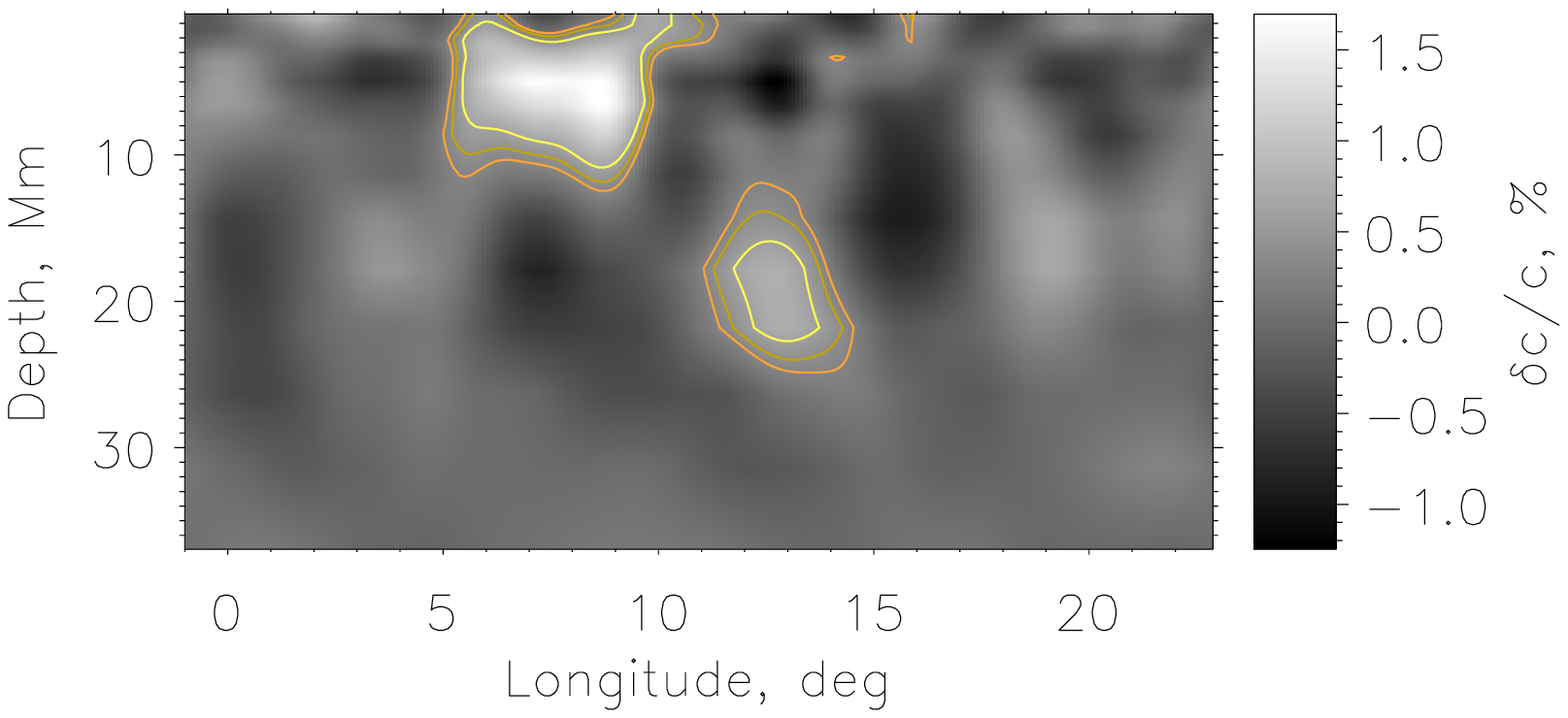} 
\includegraphics[width=6cm]{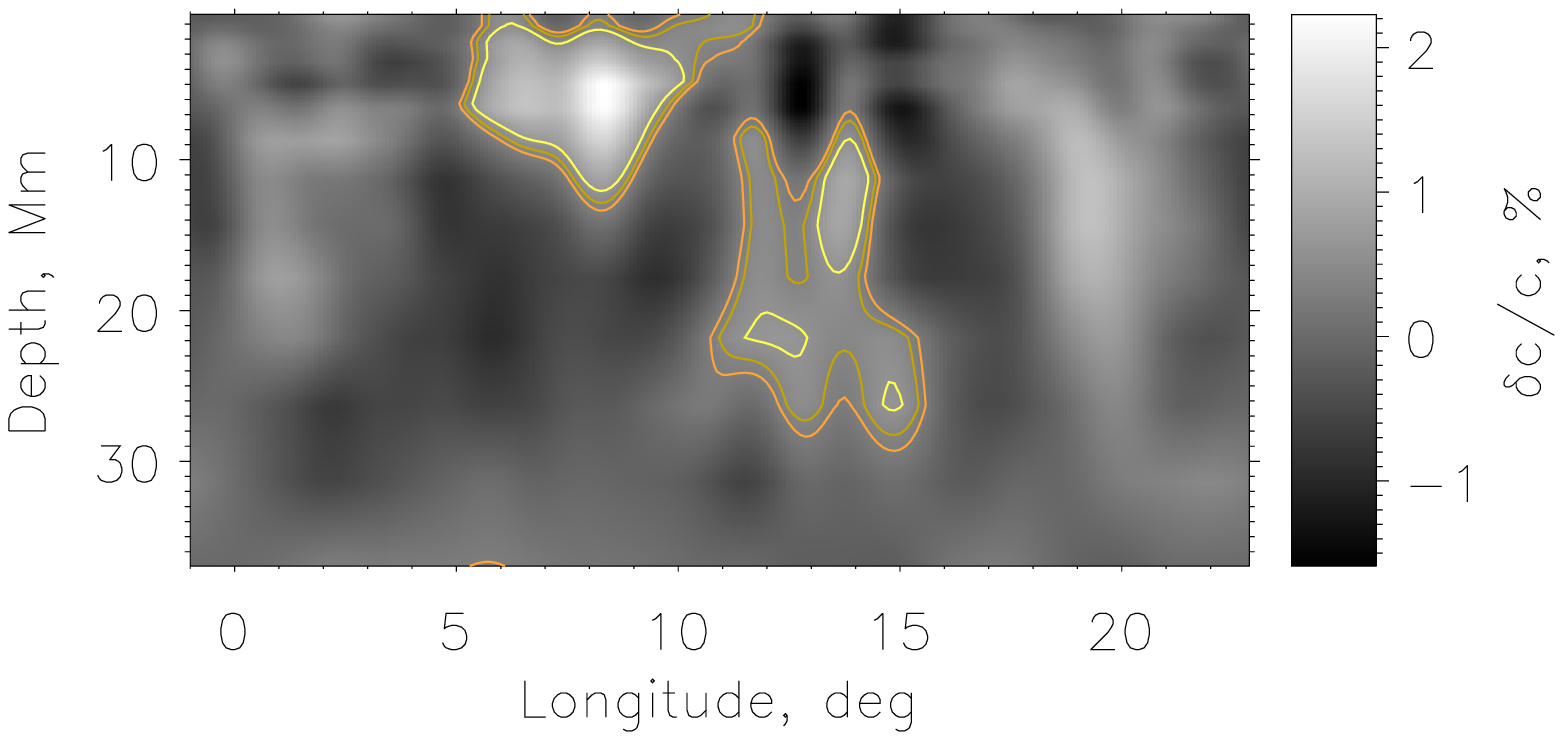} \\
\includegraphics[width=6cm]{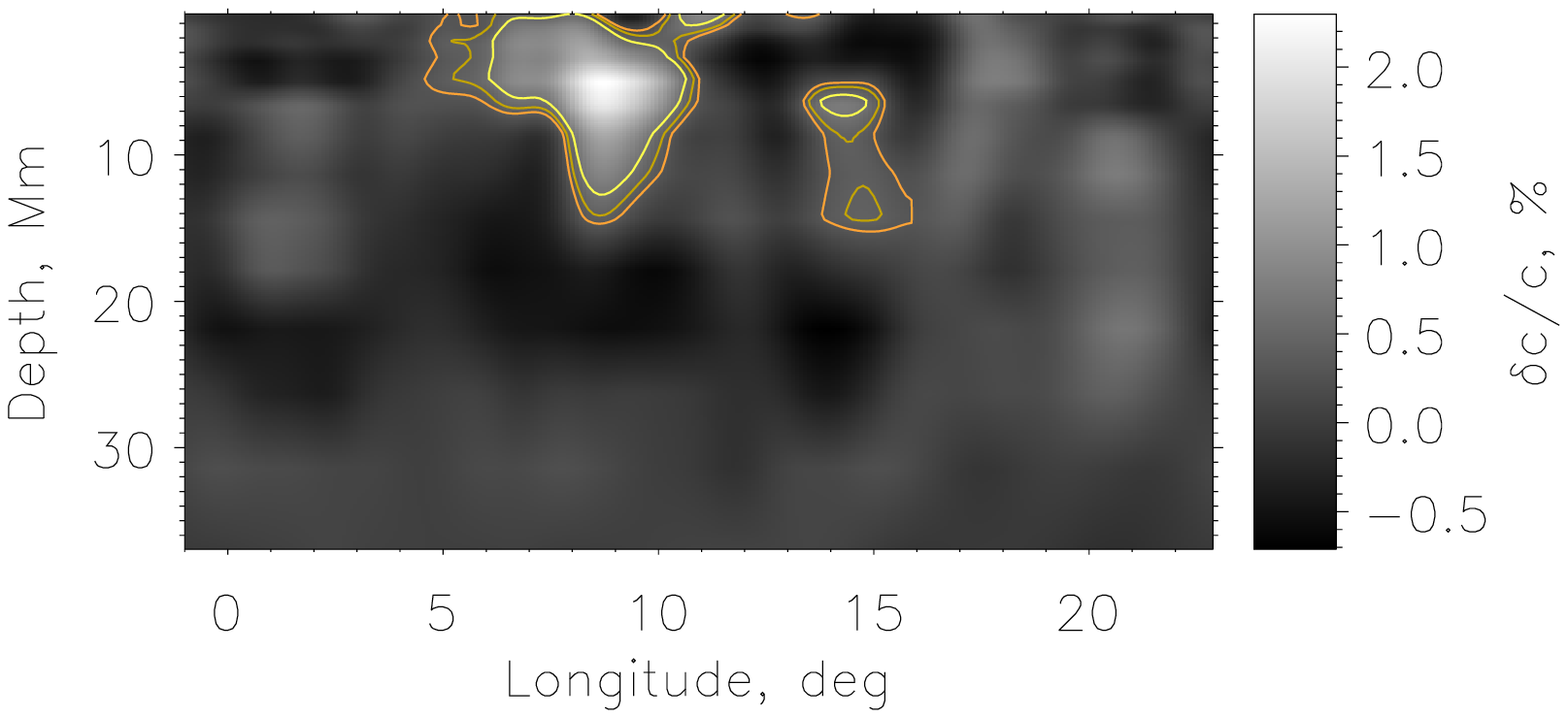} 
\includegraphics[width=6cm]{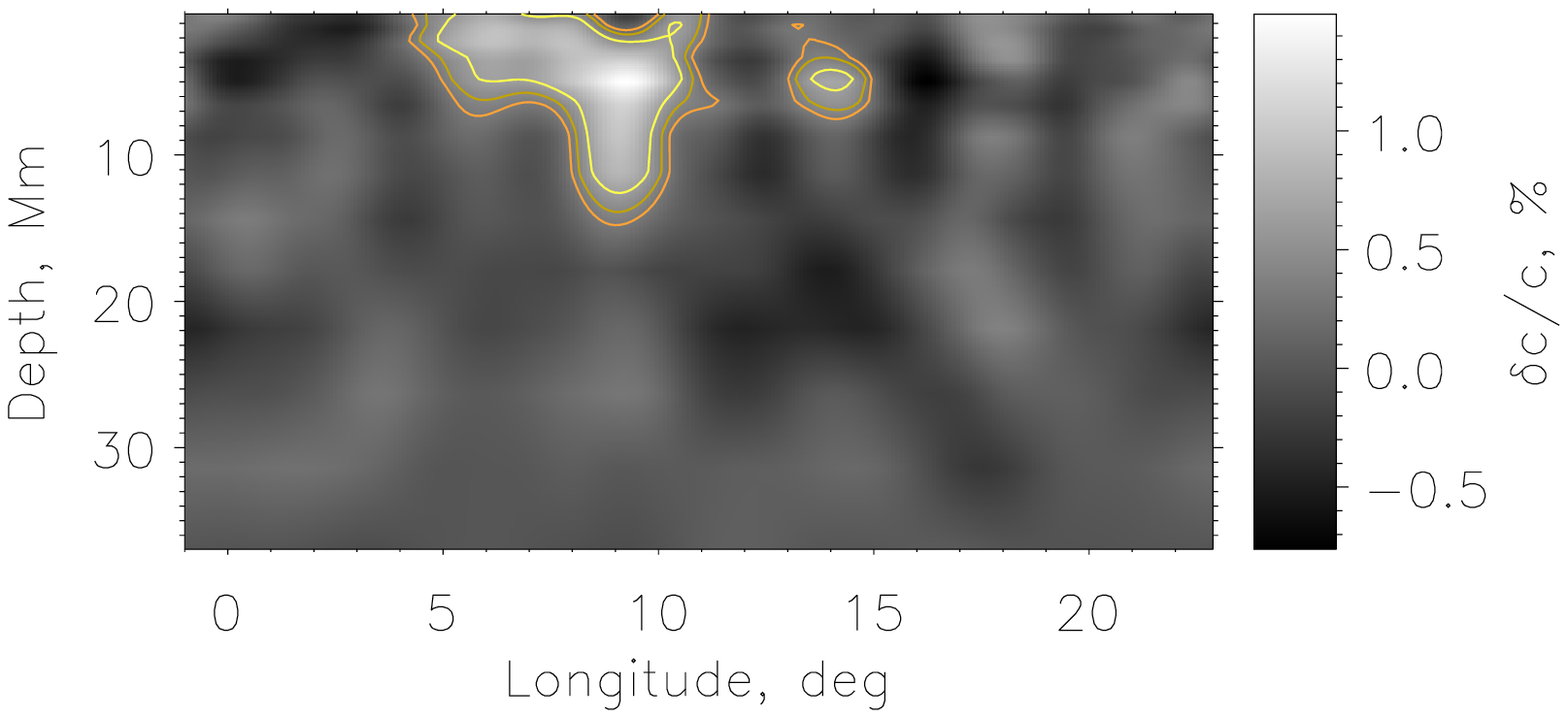} 
\caption{The magnetic flux emergence. The sound-speed inversion as a function of depth and longitude. The cut is made at $11^\circ$  latitude. 
The inversion is obtained for the 512-minute datacubes measured on 13 July 2005. The central times for the Dopplergram time series used to obtain the inversion are ({top row}): UT 04:00; 06:00; and ({bottom row}) UT: 08:00; 10:00. Contours drawn for the features located between $4^\circ$ and $16^\circ$ longitude correspond to sound-speed perturbation values of $0.5$, $0.7$ and $1.0$ per cent. 
}
\label{fig:fluxEmerge}
\end{center}
\end{figure}

\begin{figure}
\begin{center}
\includegraphics[width=6cm]{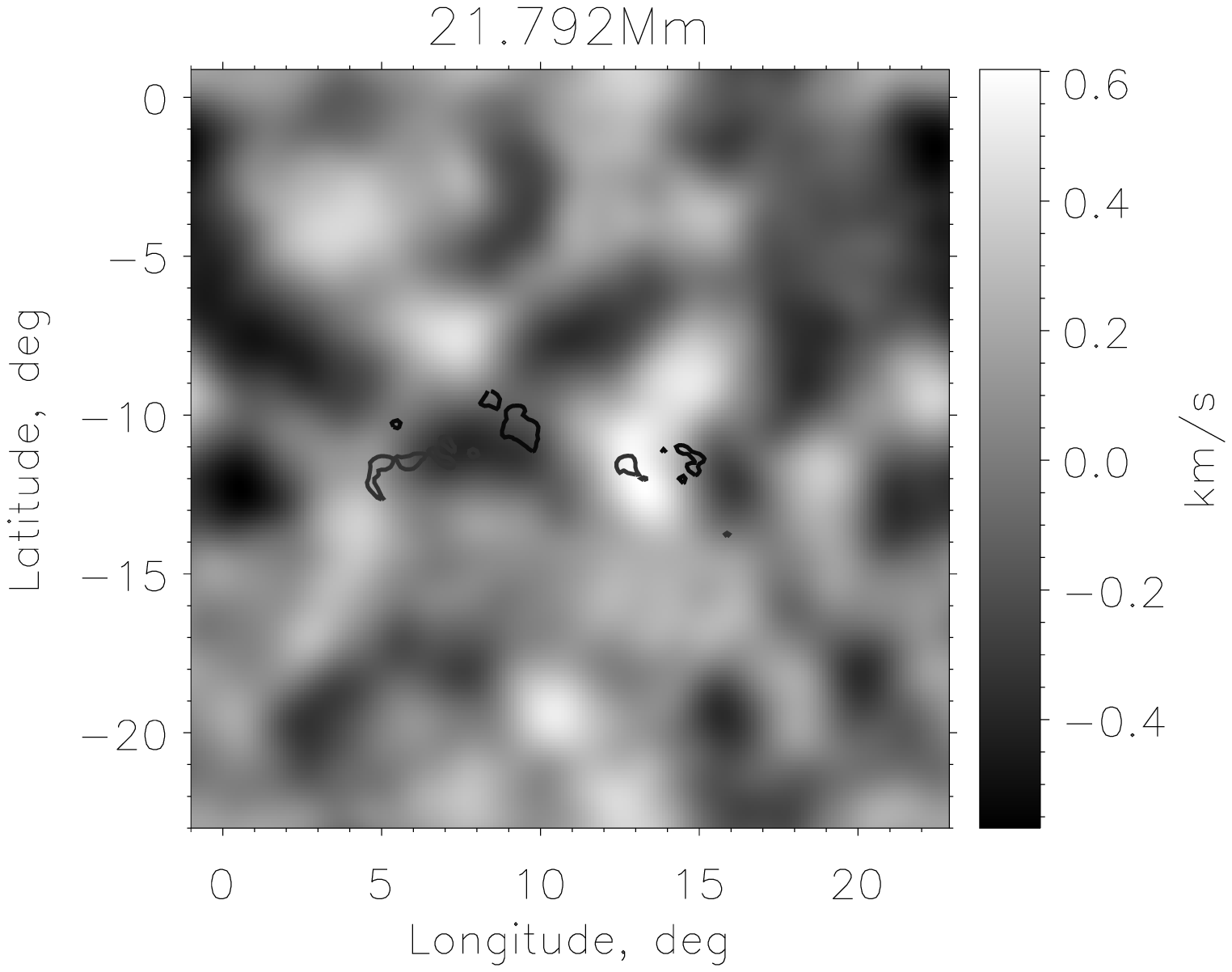} 
\includegraphics[width=6cm]{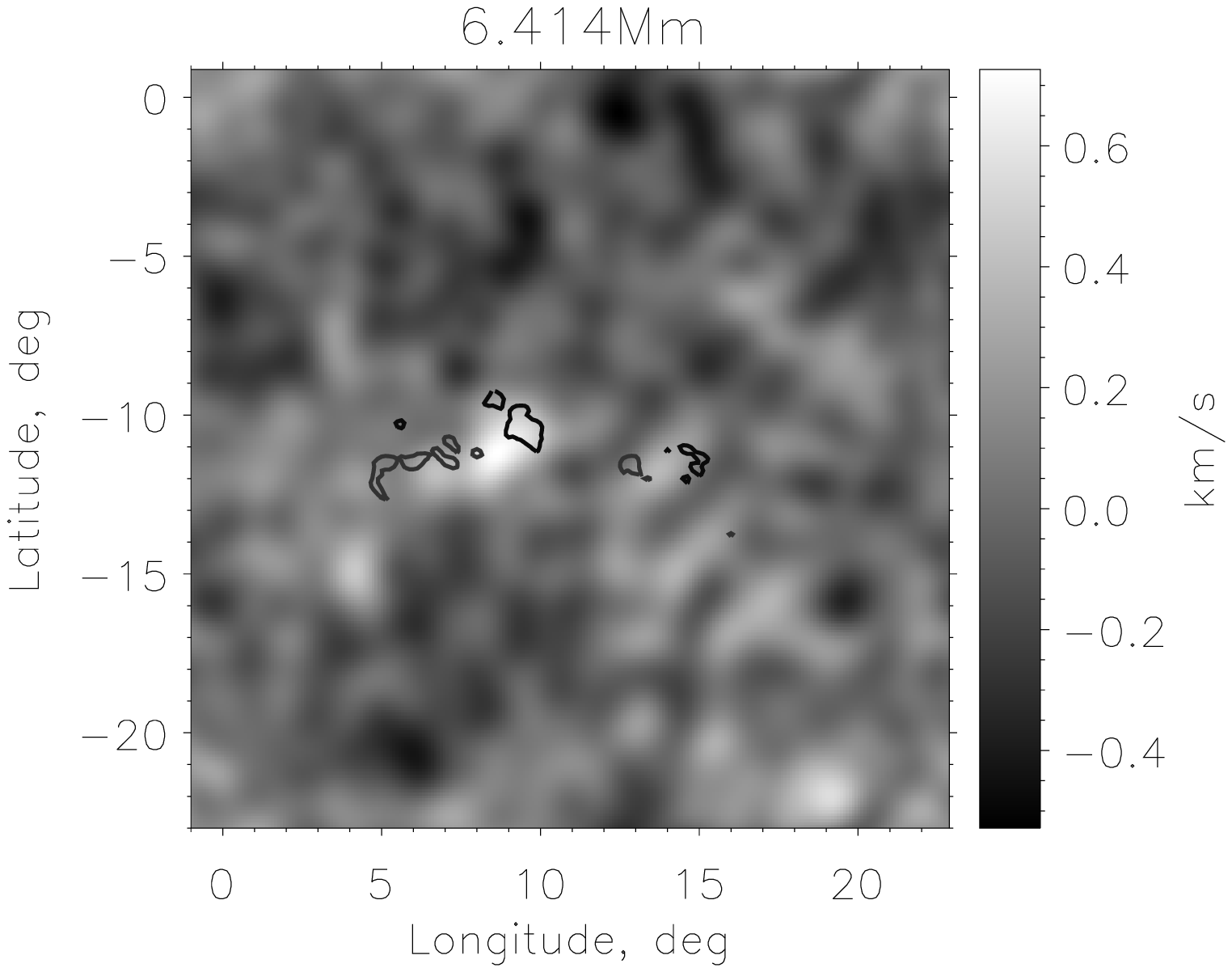} \\
\caption{{Left}: the results of the inversion at 21.77 Mm for the data cube around 13 July 2005 04:00\,UT. 
{Right}: results of the inversion at 6.41 Mm for the data cube centered around 13 July 2005 08:00\,UT. The contours in both plots are lines of $\pm 150$ Gauss from the magnetogram taken around 13 July 2005 17:35\,UT.
}
\label{fig:fluxEmergeZ}
\end{center}
\end{figure}

Three-dimensional visualisation of the sound-speed inversion for an eight-hour series centered around 04:00\,UT on 13 July 2005 is presented in Figure \ref{fig:inv3D}. One can clearly identify the subsurface tooth-like structure of the increased sound speed directly underneath the first region of surface flux activity. In addition, an object characterised by increased sound-speed  can be located deeper at the horizontal position corresponding to the point of the second flux emergence. Longitudinal cuts in depth taken at $11^\circ$ of latitude of sound-speed inversions for datacubes centered between 04:00\,UT and 10:00\,UT on 13 July 2005 are presented in Figure \ref{fig:fluxEmerge}, with depth cuts at 21.8 Mm and 6.41 Mm of inversions for series centered on 04:00\,UT and 08:00\,UT plotted in  Figure \ref{fig:fluxEmergeZ}. 

One can see the region of increased sound speed at depths of 20\,--\,25 Mm at about $14^\circ$ longitude (see Figure \ref{fig:fluxEmerge}, top left, and Figure \ref{fig:fluxEmergeZ}, left). 
This region can be tracked rising on the next inversion as shown in Figure \ref{fig:fluxEmerge} (top right panel). The regions of increased sound speed can also be seen in the following two inversions (Figure \ref{fig:fluxEmerge}, bottom panels) rising to the surface at the location of the second flux group formation. The time of the appearance on the surface (13 July 2005 UT\,06:00\,--\,14:30) corresponds to the first signs of the increased magnetic activity in the surface magnetograms. On the other hand, if the flux,  by our assumption represented in the inversions as a region of the increased sound speed, travels from deeper layers to the surface, one should be able to see the traces of this flux presence at deeper layers in the inversions of eight-hour series centered at 08:00 and 10:00, while our inversions show only localised area of sound-speed increase.  The averaged resolution kernels estimated for the depths layer around 21.79 Mm at half maximum extend from 8 to 26 Mm.

\begin{figure}
\begin{center}
\includegraphics[width=9cm, height=5.cm]{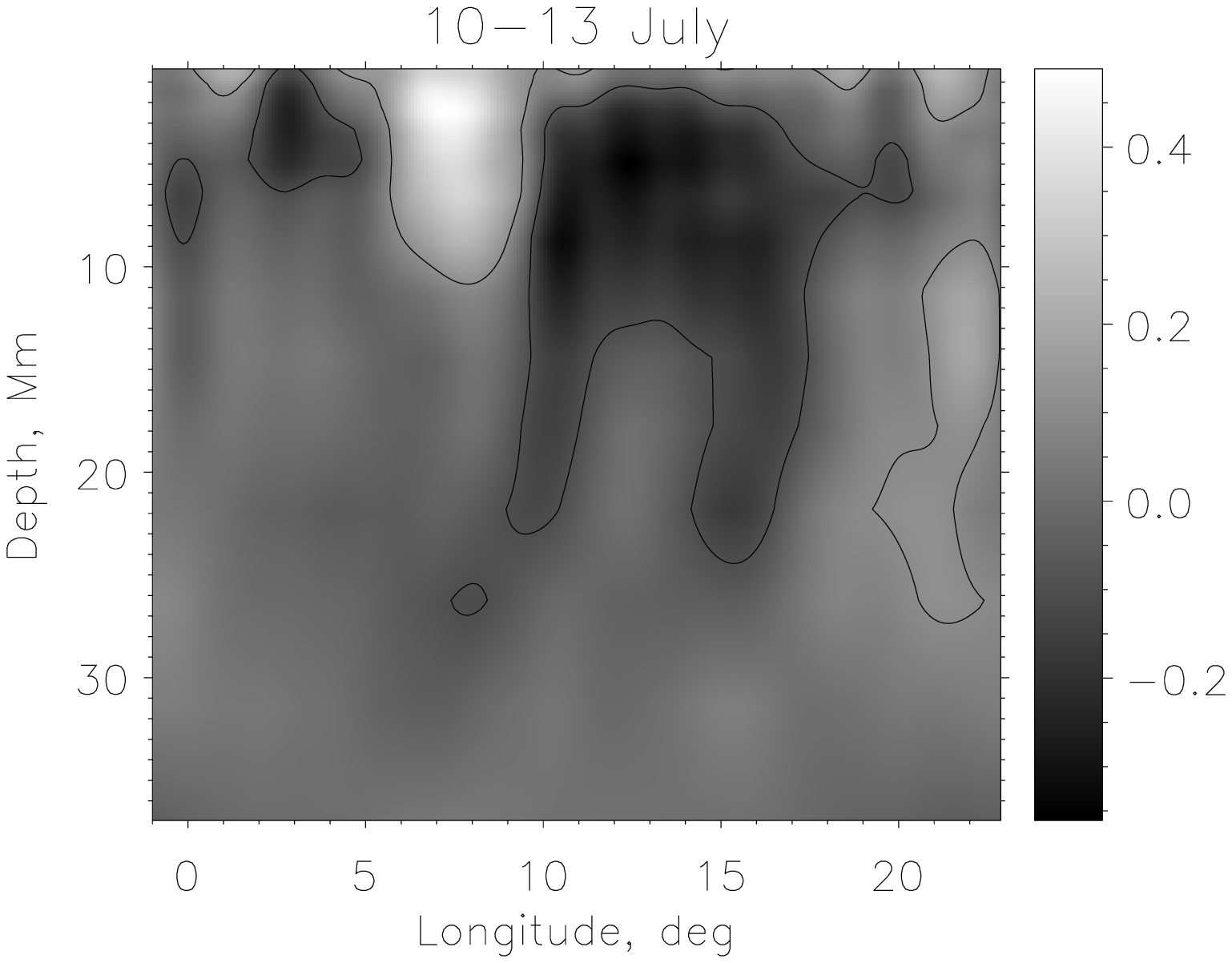} 
\\ \includegraphics[width=6cm]{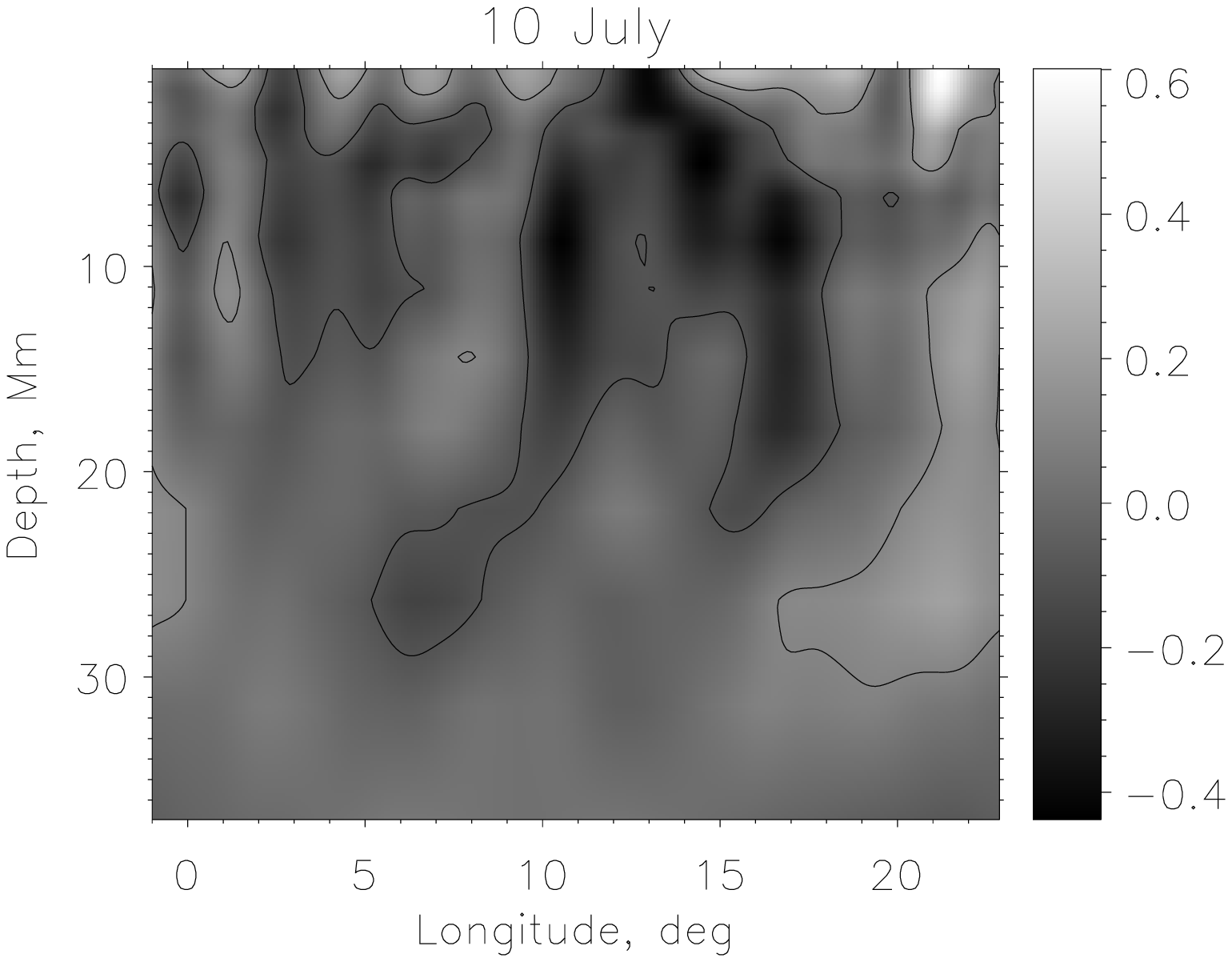} \includegraphics[width=6cm]{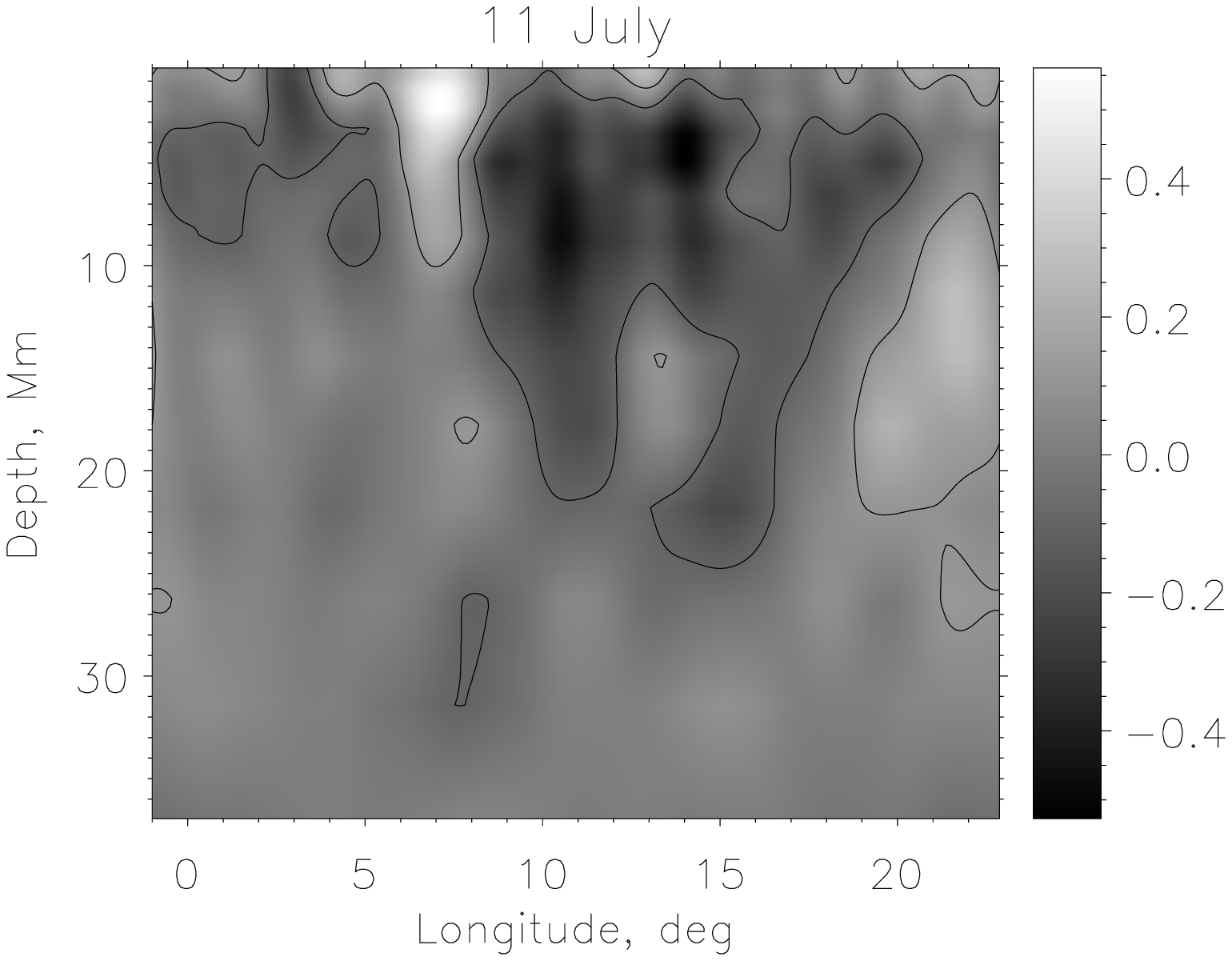} \\
 \includegraphics[width=6cm]{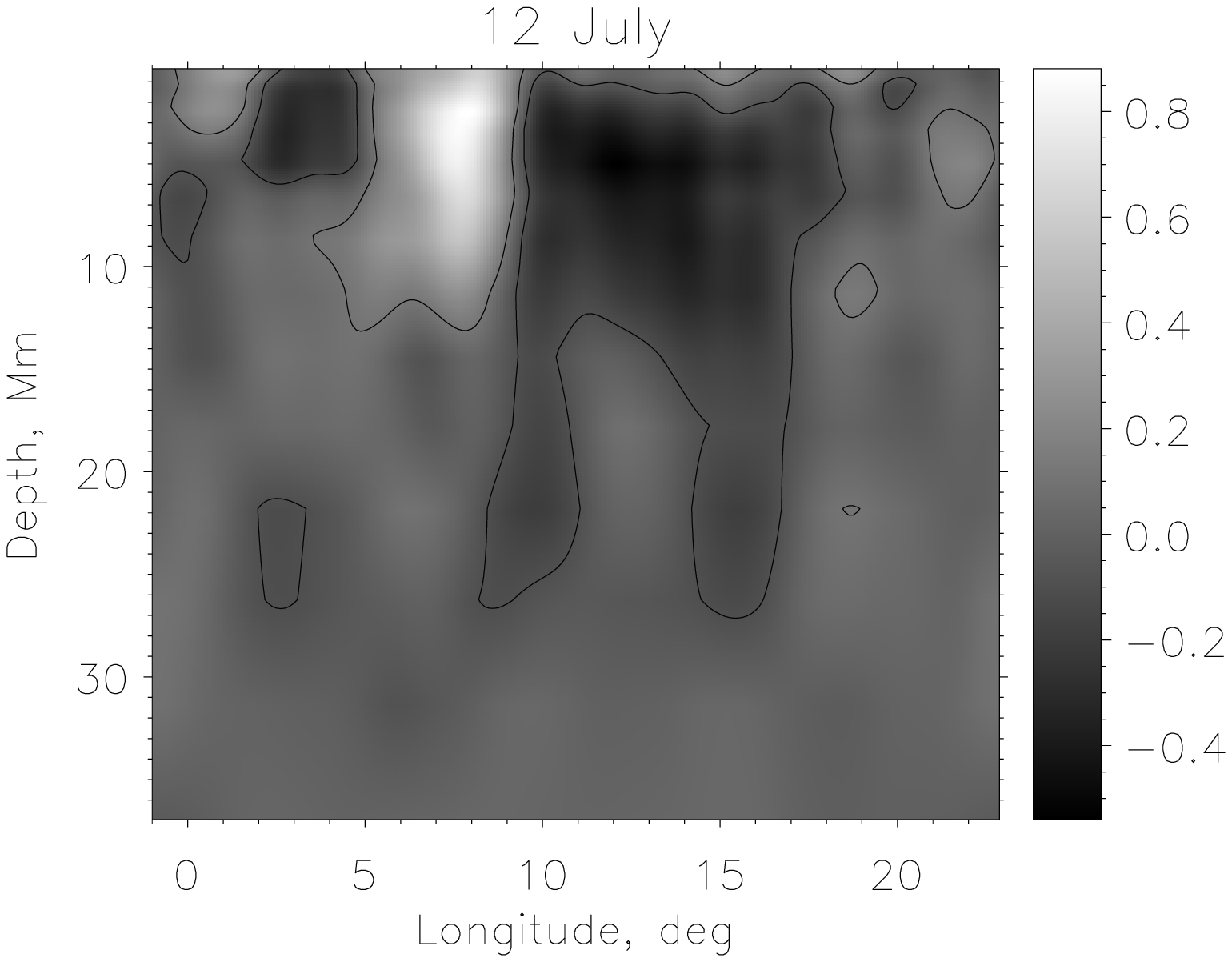}  \includegraphics[width=6cm]{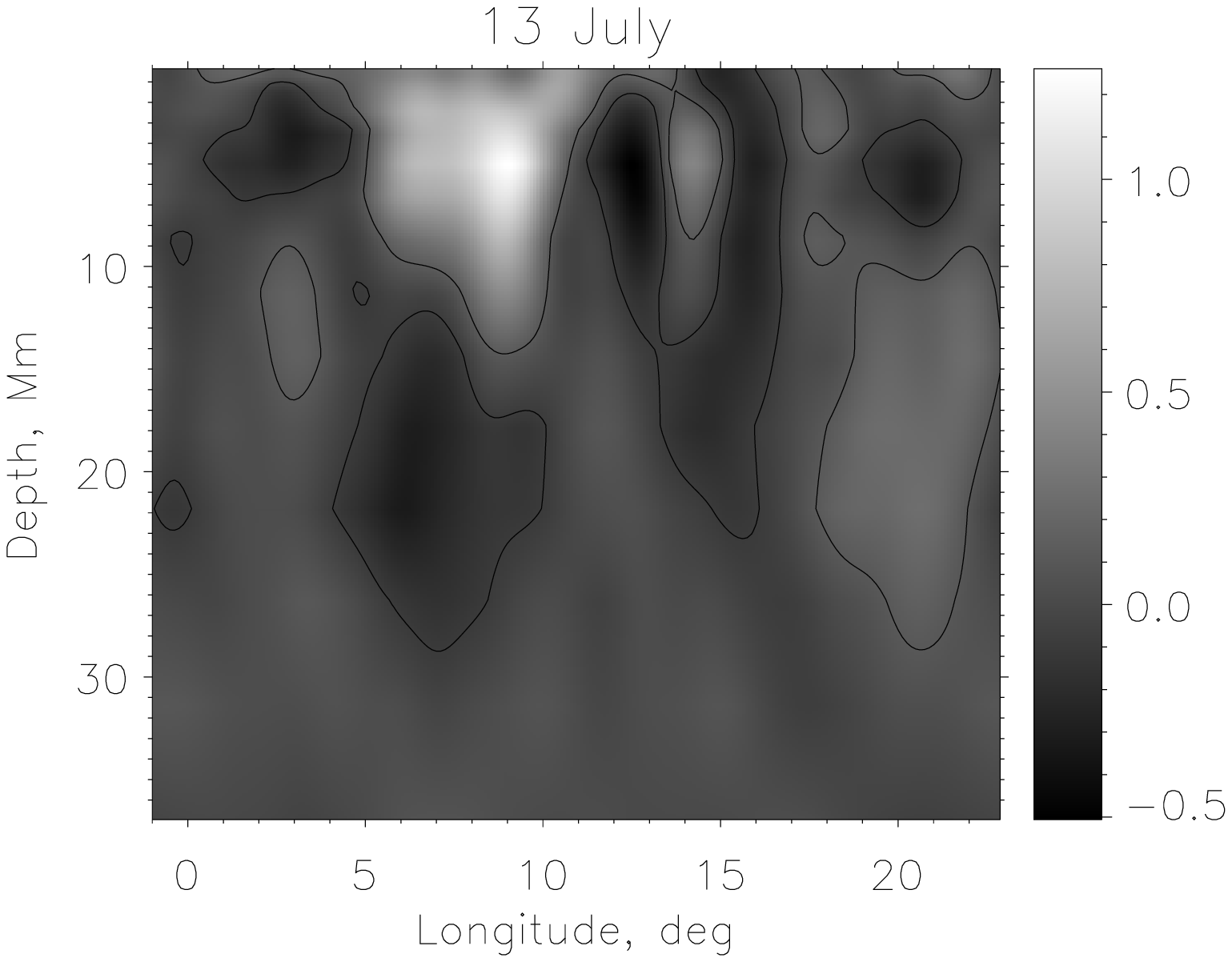} 
\caption{The mean sound-speed perturbation, averaged over all of the processed data, 44 datacubes, (top left) and the mean sound-speed perturbation for the selected dates (averages of 12 datacubes). The greyscale is $\delta c /c$ in percent. The contours correspond to sound-speed perturbation of 0.1 percent.}
\label{fig:averaged}
\end{center}
\end{figure}

In order to check if the signals corresponding to sub-surface emerging flux add up
while  the  noise  structures  cancel  out, we have averaged all of our forty-four 3D sound-speed inversion maps.  The  averaging  procedure  is
consistent  within the  linear limit used  in the inversion scheme. The depth cut at $11^\circ$ latitude of such an average is presented at the top of Figure \ref{fig:averaged}, with daily averages plotted in other panels.
One can clearly see the region of the increased sound speed corresponding to the first flux location. Interestingly, one can also observe a well defined region of the decreased sound speed located beneath and to the West of the first flux region with a column of negative perturbation corresponding to the location of the second flux emergence. The structure is stable with respect to changes in the pre-processing (filtering) scheme and inversion parameters. It is also clearly present in the averages of the datacubes obtained on any of the days considered here, from 10 to 12 July 2005 presented in Figure \ref{fig:averaged}. However, the shape of this decreased sound speed cannot be observed in individual eight-hour inversions, possibly owing to turbulent activity. 

We are not certain how to interpret this region of negative sound-speed perturbation but presumably it is a region of cooler plasma. Under certain conditions such a signal could arise from a monolithic flux tube originating in the deeper interior, 
with internal density exceeding the matter density outside the tube. Alternatively, 
such an effect can be seen in the scenario described by \inlinecite{kitch} in which a temperature instability 
leads to the formation of an active region due to eddy diffusivity and magnetic tension. If the former explanation is the correct one, then it is tempting to identify such emerging cool flux tubes with the flux tubes postulated by \inlinecite{SZ99} to explain transient darkenings they observed in continuum intensity.


Thus, we can hypothesize that the sound-speed perturbation seen in Figure \ref{fig:fluxEmerge} is related to the 
emergence of a hot packet of plasma in the cold plasma column seen in Figure \ref{fig:averaged}. In such circumstances, the 
absence of a positive perturbation column in the bottom two images of Figure \ref{fig:fluxEmerge} may be due to 
the dynamic nature of the region considered, wherein the body of increased sound speed preceded and followed by 
regions of decreased sound speed may cancel each other in our inversion. 
We note that the inversions of the interior sound speed averaged over eight datacubes obtained on 13 July 2005 (Figure \ref{fig:averaged}, bottom right panel) point to the presence of a decreased sound-speed structure column at the location of the second flux emergence with a positive perturbation directly below the surface.
The latter feature can be expected given that these data are for a period after the flux emergence at the surface has become well established.
So, if we assume that the region of the increased sound speed 
corresponds to the emerging magnetic flux and compare the sound-speed perturbation detected at the depth of 21.7 Mm on 13 July 2005 04:00 and appearing on the surface around 10:00\,UT, this can provide an estimate of the speed of 
emergence at about 1.0 ${\rm km\ s^{-1}}$. This estimate is slightly lower than the speed reported by \cite{Kosovichev2000}, 
but definitely higher than the velocities of a diamagnetic transport in near-surface layers estimated in \cite{Kriv05}. This can be an indication of some additional mechanisms affecting the flux emergence.

In our solutions, the errors due to realisation noise are less than 100 ${\rm m\,s^{-1}}$: this is 
estimated by processing the 24 quiet-Sun datacubes and  applying the approach described in \cite{jensen01, GB04, Cv05}. 

\section{Conclusions}
\label{Sec:Conclusions}
We have investigated the emergence of the Active Region NOAA 10790. The region has a complex magnetic structure and can be viewed as an amalgamation of two magnetic region as described in Section \ref{sec:DisData}. 
We see shallow regions of increased sound speed at the location of increased magnetic activity, with regions becoming deeper at the locations of sunspot pores. It appears that a subsurface sound-speed perturbation at depth is associated with the emergence on the surface of a sunspot and sunspot pores. 
We observe an object with increased sound speed which may be related to the subsurface flux prior to its emergence in the photosphere at the second location. 
We also observe a loop-like structure of the decreased sound-speed perturbation located below the investigated region, which is present on the averages of the inversions. It would be interesting to see if similar structures characterised by a drop in sound speed can be detected in other instances of emerging active regions.

Based on our results, one can speculate that magnetised regions without any sunspots or sunspot pores lead to increased subsurface sound speed at shallow depths but no perturbation detected at greater depths. On the other hand, the areas associated with sunspot appearance have increased finger-like sound-speed perturbations which extend to depths of 10\,--\,20 Mm. Then by tracking the times of the emergence of magnetic flux in the subsurface layers of the Sun one can estimate the speed at which flux emerges to be about $1 {\rm\, km \, s^{-1}}$. 

We also note that the precise timing of the signature of emergence in the seismic imaging is very limited by the temporal resolution of  the sound-speed inversions presented here, since each set is a representative of eight hours of solar oscillations. 

In view of the dynamo models discussed in Section \ref{sec:Intro} (\opencite{NC02}, \opencite{Kriv05}), it could be interesting to compare such results for a variety of emerging active regions at different latitudes and stages of the solar cycle. 

\acknowledgements
We would like to thank an anonymous referee for the valuable suggestions that have greatly improved this paper. We also acknowledge the support from UK Science and Technology Facilities Council grant number PP/C502914/1.
\bibliographystyle{spr-mp-sola}

\end{article} 
\end{document}